\documentclass[mnsc,nonblindrev]{informs3_nojournal} 

\OneAndAHalfSpacedXI 

\usepackage{endnotes}
\let\footnote=\endnote
\let\enotesize=\normalsize
\usepackage{natbib}
 \bibpunct[, ]{(}{)}{,}{a}{}{,}%
 \def\bibfont{\small}%

\usepackage{graphicx}
\usepackage{multirow}
\usepackage{subcaption}
\TheoremsNumberedThrough     
\ECRepeatTheorems

\EquationsNumberedThrough    

\begin{document}


\RUNAUTHOR{Bergquist and Elmachtoub}

\RUNTITLE{Static Pricing Guarantees for Queueing Systems}

\TITLE{Static Pricing Guarantees for Queueing Systems}

\ARTICLEAUTHORS{%
\AUTHOR{Jacob Bergquist}
\AFF{Department of Industrial Engineering and Operations Research, Columbia University, New York,
NY, 10027, \EMAIL{jacob.bergquist@columbia.edu}
} 
\AUTHOR{Adam N. Elmachtoub}
\AFF{Department of Industrial Engineering and Operations Research \& Data Science Institute, Columbia University, New York,
NY, 10027, 
 \EMAIL{adam@ieor.columbia.edu}}
}

\ABSTRACT{%
We consider a general queueing system with price-sensitive customers in which the service provider seeks to balance two objectives, maximizing the average revenue rate and minimizing the average queue length. Customers arrive according to a Poisson process, observe an offered price, and decide to join the queue if their valuation exceeds the price. The queue is operated first-in first-out, can have multiple servers, and the service times are exponential. Our model represents applications in areas like  make-to-order manufacturing, cloud computing, and food delivery.

The optimal solution for our model is dynamic; the price changes as the state of the system changes. However, such dynamic pricing policies may be undesirable  for a variety of reasons. In this work, we provide non-asymptotic performance guarantees for a simple and natural class of static pricing policies which charge a fixed price up to a certain occupancy threshold and then allow no more customers into the system. Despite the mixed-sign objective, we are able to show our policy can guarantee a constant fraction of the optimal dynamic pricing policy in the worst-case. We also show that our policy yields a family of bi-criteria approximations that simultaneously guarantee a constant fraction of the optimal revenue with at most a constant factor increase in expected queue length.  For instance, our policy for the M/M/1 setting can be set so that its worst-case guarantees is at least 50, 66, 75, or 80\% of the optimal revenue and at most a 0, 16, 54, or 100\% increase in the optimal queue length, respectively. We also provide guarantees for settings with multiple servers as well as the expected sojourn time objective. In a large simulation, we show that our class of policies is at most 4\% sub-optimal on average. 
}%




\maketitle


\section{Introduction}

There are many business applications which can be naturally modeled as queueing systems with prices. One prototypical example is a firm which provides make-to-order goods such as custom electronics or vehicles. In this application, customers approach the firm with a request after which the firm quotes a price. If the price is acceptable, the customer submits their order and joins a virtual queue. Another important modern application is in cloud computing. Here, the arrival process of customers seeking cloud computing resources is modulated by the price posted by the service provider, and customers accepting the price will put their jobs in queue to await service. The food delivery industry is also based on customers deciding whether or not to order based on delivery fees, and customers join a virtual queue to wait for their food if they make a transaction. In such applications, a tradeoff often exists between the congestion the service provider allows and the revenue they earn. In the short term, one may be able to earn extra revenue by allowing a lot of congestion, but large queues may have a long term negative effect on demand. To manage this tradeoff, one approach is to price \textit{dynamically}, changing the price as the state of the system changes. Dynamic pricing allows the service provider to control the congestion in the system without sacrificing much revenue: as the number of customers in the queue increases, the service provider can charge higher prices to reduce the rate of purchasing customers that need to be serviced. Indeed, the optimal policy for this class of models involves dynamic pricing.

In practice, however, there are many downsides to dynamic pricing. The strategy may be unappealing to customers and appear unfair \citep{cohen2022price}. In response, customers might begin to exhibit strategic behavior such as delaying their purchase, which may not be accounted for in the model. Implementing a dynamic pricing strategy also requires increased operational complexity to communicate and update prices. Moreover, a  reliable demand model is needed to ensure small price changes are indeed causing intended changes in demand. 
To address these issues, \textit{static} pricing may be an attractive alternative. Static pricing is advantageous not only because it provides transparency for the customer, it is also simpler and more tractable for service providers to implement. However, when we restrict ourselves to static pricing policies, some loss of revenue or increased congestion may have to be endured. While it is understood that static policies are optimal in the fluid limit of many stochastic models, their theoretical performance is less understood when service and arrival rates are finite. In this work, we quantify the strength of static pricing in the worst-case by furnishing universal, non-asymptotic guarantees  relative to the optimal dynamic pricing policy.

We consider a generalization of the classic $M/M/C$ queueing model. We assume that customers arrive according to a Poisson process, service times are exponentially distributed, and customers' valuation distributions are regular. The service provider sets a price that can change with the state, and customers purchase the service if their valuation exceeds the price. The class of regular distributions is standard to consider in revenue management and auction theory, and includes the uniform, exponential, Gaussian, and some heavy tail distributions. The service provider seeks to optimize two objectives, maximizing revenue while minimizing congestion. To reduce this multi-objective problem to a single-objective one, we assume the service provider seeks to maximize their expected revenue rate minus some constant times the average number of customers in the system. This constant captures the desired penalty the service provider wishes to associate with congestion in the system. For example, in the context of cloud computing, the congestion penalty may come from the cost of memory on the servers which hold the jobs in queue and from the reputational loss incurred when delays are too long. We also consider a related model in which the congestion penalty associated with a policy is proportional to the long-run average sojourn time under that policy. 

The static pricing policies we consider use a single price, and stop selling when the number of customers in the queue reaches a specified threshold. These threshold-static pricing strategies are simple to implement and allow us to directly trade off revenue and congestion via the threshold parameter. Note that for static policies which do not have a threshold, we show the performance can be arbitrarily poor. We emphasize that our guarantees are universal, meaning that they do not depend on the instance and are not asymptotic in nature. Our proofs are by construction, and we use the static price that results in the same expected arrival rate as the optimal policy. This construction has two key advantages: \textit{(i)} the revenue rate when the state is below the threshold is always at least as good as the optimal policy due to an application of Jensen's inequality \textit{(ii)} we can directly relate our policy to the expected queue length of the optimal policy via Little's law. However, this use of the optimal policy's average arrival rate (denoted $\tilde{\lambda}$) is purely a proof device; in practice, one would not need to compute $\tilde{\lambda}$ or find the optimal dynamic policy. Instead, a decision-maker can search directly over the two parameters (the price and cutoff threshold) to implement a static policy.

We now summarize the main contributions of the paper.

\begin{itemize} 
\item When there is a single server, which corresponds to the classic M/M/1 queue, Theorem \ref{singleServerProfit} shows that a static pricing policy can always attain at least half of the optimal value of the optimal dynamic pricing policy. We note that the objective is mixed-sign, and it is notoriously challenging to generate approximation guarantees for such settings since the optimal value can be arbitrarily close to zero. 

   \item For our static policy, we can obtain a family of bi-criteria approximations via Theorem \ref{StaticGuarantee} that simultaneously guarantee a constant fraction of the optimal revenue with at most a constant factor increase in expected queue length.  For instance, our policy for the M/M/1 setting can be set so that its worst-case guarantees is at least 50, 66, 75, or 80\% of the optimal revenue and at most a 0, 16, 54, or 100\% increase in the optimal queue length, respectively. Moreover, we provide a class of instances proving the tightness of our analysis.

 \item In the multi-server case, Theorem \ref{multiServerProfit} establishes that for a system with $C$ servers, a static pricing policy can always attain at least a $1-\frac{\frac{1}{C!}(C^C)}{\sum_{n=0}^C\frac{1}{n!}C^n}$ fraction of the multi-objective optimal value. Thus, the advantage of dynamic pricing over static pricing vanishes as the number of servers increase: with $10$ servers, our guarantee is at least $78.5\%$.  
 \item Theorem \ref{StaticGuaranteeMultiserverWithK} furnishes bi-criteria approximations on the revenue and congestion objectives as a function of the cutoff threshold (the maximum number of customers allowed in the system). For instance, when there are ten servers, static pricing can guarantee at least 90\% of the revenue with at most a 20\% increase in queue length. Service providers can use this result to understand the benefit of additional servers and changing the cutoff threshold in the context of managing revenue and congestion with static pricing. 
      \item Theorem \ref{SojournGuarantee} extends our results for a related model  in which we penalize long-run average sojourn times instead of average occupancy to measure congestion. In this case, we give bi-criteria guarantees for the revenue and congestion objectives.   
    \item We report the results of numerical experiments over a wide test bed demonstrating the empirical performance of our policies. These experiments illustrate the high performance of static pricing policies: in the worst-case instance we found, the optimal static pricing policy is able to recover 89.70\% of the optimal multi-objective value. We show that static policies are at most 4\% sub-optimal on average.

\end{itemize}

\subsection{Literature Review} 
We break the related literature down along three different dimensions: \textit{(i)} whether pricing is static or dynamic, \textit{(ii)} whether the analysis is asymptotic or exact (non-asymptotic), and \textit{(iii)} the manner in which the state of the queue impacts the arrival process. Along the last dimension, the literature has primarily considered three different possibilities (in order of their first appearance):
\begin{enumerate}
    \item The state of the queue does not directly impact the arrival process, which we call the Low queue since it was first introduced in \cite{10.2307/169504}.
    \item Observable queues, in which customers observe the state of the queue and use this information to inform their decisions.
    \item Unobservable queues, in which customers do not observe the state of the queue but have an expectation based on past experience of what their waiting time will be.
\end{enumerate}
In this work, we provide exact (non-asymptotic) analysis on the trade-off between static and dynamic pricing for Low queues. 

To our knowledge, the model we consider was first introduced in \cite{10.2307/169504} and followed up with \cite{5391267}, works in which the author studies the structure of the optimal dynamic pricing policy and gives an algorithm to compute it assuming either a finite or compact set of possible prices, respectively. Our work uses different objective functions motivated by our applications and does not require compactness of the possible prices.  The Low queue, albeit restricted to a single-server, was again considered in \cite{LATOUCHE1980203}. They consider two-price policies and, for a given pair of prices, compute the optimal policy under two different criteria: either maximizing revenue rate for a given upper bound on congestion, or minimizing congestion for a given lower bound on revenue rate. This contrasts with our approach in which we optimize a linear combination of the two objectives and allow a different price for each state in the dynamic policy. More recently, \cite{MAOUI2009912} considers the problem of optimizing the static price in the single-server instance of our model. However, their work is focused primarily on identifying static pricing policies and studying the structure of the optimal policies; our aim in the present work is to relate the performance of static and dynamic policies. Another work which extends Low's model is \cite{doi:10.1287/mnsc.1060.0587}, which considers the case when the service rate is variable and can be picked by the service provider. They show numerically that, in their setting, dynamic pricing has significant gains over static pricing. Recently, \cite{vaze2022nonasymptotic} considers a model closely related to the Low queue, but where the service rates (instead of the arrival rates) are taken as the decision variables. 

We now discuss works which consider observable queues. The foundational and seminal work in this line is  \cite{10.2307/1909200}, which introduced observable queues and analyzed them under the assumptions of deterministic customer valuations and static pricing. The most salient point emerging from their exact analysis is that the static price which maximizes social welfare is strictly smaller than the revenue-maximizing price. 
In \cite{doi:10.1080/07408170108936878}, the authors generalize the observable model in \cite{10.2307/1909200} (and their work in \cite{articleChenFrank2}) to allow for state-dependent pricing. Their work shows that a threshold policy is optimal but does not give explicit expressions for the threshold. The work of \cite{borgs_chayes_doroudi_harchol-balter_xu_2014} provides explicit expressions for the optimal threshold in terms of the Lambert-W function.  \cite{GILBOAFREEDMAN2014217} considers the gain of dynamic over static control of a queueing system under an extension of Naor's model where some customers are assumed to cooperate with the service provider. 
\cite{doi:10.1287/opre.2017.1668} analyze the value of dynamic pricing in the asymptotic regime, where the arrival and service rates are taken to infinity. They show that dynamic pricing has a significant effect on reducing the impact of stochasticity on accrued revenue, though most of the benefit can be attained by simple two-price policies. Examples of related work in this line are \cite{842140}, \cite{articleyoon}, \cite{doi:10.1287/mnsc.49.8.1018.16402}, \cite{doi:10.1287/opre.1060.0305}, and \cite{articleCil}. 

We now discuss a few works which assume the queue is unobservable. The earliest study of this case is \cite{10.2307/1913415}, where they show that the revenue maximizing and social welfare maximizing prices are the same. 
Another work in the unobservable case is \cite{10.2307/171046} which considers a multi-class model and provides static pricing mechanisms which maximize total social welfare. \cite{RePEc:inm:ormsom:v:13:y:2011:i:2:p:244-260} carries out an exact analysis assuming an unobservable queue with static prices to identify when a subscription-based pricing scheme (in which users pay a monthly fee for an arbitrary number of uses) may outperform a per-use approach. Again assuming static pricing, \cite{doi:10.1287/msom.2014.0479} analyzes the loss of optimality that must be suffered when a price must be set without knowledge of the arrival rate.

The preceding works and the literature more generally have mostly dealt with the case where the queue is either completely observed or totally opaque to the customer; another avenue is the question of interpolating between these two extremes by investigating optimal signaling. This question has been addressed in a few works, including \cite{HASSIN2017436} and \cite{doi:10.1287/opre.2018.1819}. In these works, the authors show that with respect to a certain objective, the optimal signaling policy falls in the middle of the two extremes. 

The foregoing works are all unified in that the underlying queueing model is essentially an $M/M/C$ queue. We continue by reviewing works which consider models slightly further afield.  \cite{doi:10.1287/opre.2020.2054,elmachtoub2023power} provides universal performance guarantees for static pricing policies when the model is an Erlang loss system (which has no congestion objective), and describe a real-world implementation in \cite{besbes2020pricing}. 
\cite{4604751} introduce a  stylized model involving pricing and queues that is tailored for telecommunication applications and furnish a lower bound on the ratio between the revenue earned from their simple pricing rule and the maximum possible, which they denote as the ``price of simplicity". Their model assumes no disutility is incurred for long delays in service. \cite{6354274} shows that when such disutility is assumed to be present, the ``price of simplicity" can be very high. Since in typical applications high delays are undesirable, this insight demonstrates the importance of considering congestion in these models and motivates our centering of it in the present work. Examples of other important works which analyze pricing in queueing models that are tailored for more specific applications are \cite{doi:10.1287/mnsc.1060.0651}, \cite{6671562}, \cite{articleWasherholeJost}, \cite{10.1145/2764468.2764527},  \cite{10.1145/3033274.3085099}, \cite{doi:10.1287/mnsc.2022.4474}, and \cite{benjaafar2023pricing}. Finally, \cite{elmachtoub2023simple} show analogous bi-criteria approximations to ours but for a completely different problem regarding joint pricing and inventory control.


\subsection{Organization}
Our work is organized as follows. In Section \ref{Model+Prelims}, we provide the general model and describe our static pricing policies. In Section \ref{StaticCutoffSection}, we prove our revenue and cost rate approximation guarantees for static pricing in the single-server case.  We extend the results to the multi-server case in Section \ref{multiServerSection} and to the expected sojourn times as the congestion objective in Section \ref{sojournModel}. Numerical experiments are provided in Section \ref{experimentsSection}.  Finally, we conclude our paper and offer some future directions in Section~\ref{conclusion}.

\section{Model}\label{Model+Prelims}

 We consider a model  where a service provider seeks to optimize a linear combination of the expected revenue and the expected congestion of the system. There are $C$ servers that serve price-sensitive customers in a first-come, first-serve manner. Customers arrive according to a Poisson process with rate $\Lambda > 0$. Each customer has a valuation drawn i.i.d. from a distribution $F$. We make the standard assumption in the revenue management literature that $F$ is a regular distribution (also known as Myerson's regularity, see \cite{doi:10.1287/moor.6.1.58}). When a customer arrives, the provider offers service at some price $p$ which may depend on the state of the system; the customer decides to join the queue if their valuation is at least $p$. This gives rise to the \textit{effective} arrival rate of customers, which we denote by $\lambda(p):=\Lambda (1-F(p))$. 
We assume that there is a one-to-one correspondence between prices $p$ and effective arrival rates $\lambda$ so that $\lambda(p)$ has a unique inverse, denoted by $p(\lambda)$. Thus we can equivalently view the effective arrival rates $\lambda$ as the decision variables, which simplifies our analysis. By our assumption that $F$ is regular, we have that $\lambda p(\lambda)$ is concave  \citep{Ewerhart2013RegularTD}. 

Each customer has a service time which follows an exponential distribution with mean $1/\mu$. We assume that the service times are i.i.d. across customers and independent of customer valuations.
 The firm incurs a congestion penalty of $c$ per time unit per customer in the system. The value $c$ can be viewed as a parameter which captures the magnitude of the penalty the service provider wishes to associate with congestion. This penalty structure amounts to penalizing a policy proportionally to the expected number of customers in the system. A related quantity, the expected sojourn time of a customer in steady-state, is also a natural choice for a penalty in many applications. We  consider this quantity as an extension of our work in Section \ref{sojournModel}.

 The parameters for our model depend upon the units we choose for currency and time. By picking the units in a certain way, we can transform an arbitrary instance it into one with $\mu=c=1$ without impacting the approximation ratios of interest. Specifically, given an arbitrary instance, we can translate into an instance with $\mu=c=1$ by measuring all monetary values in units of $c$ and time in units of $\mu$. For example, if $\Lambda = 2 \text{ per hour}$, $\mu = \text{ 1/2 per hour}$, $c = \$50 \text{ per hour}$, and $\lambda(p) = 2-p/100 \text{ per hour}$ (where $p$ is in dollars), we can instead consider the instance with $\Lambda' = 4 \text{ per half hour}$, $\mu' = 1 \text{ per half hour}$, $c'= 1 \text{ hundred dollars per half hour}$, and $\lambda'(p) = 4-2p \text{ per half hour}$, (where $p$ is now measured in hundreds of dollars). In general, the ``nominal" revenue, cost, and profit rates garnered by a policy change by a factor of $\frac{1}{\mu c}$ when making this translation. In the performance ratios that the present work is concerned with, this factor is present in both the numerator and denominator and thus cancels out. Hence, we assume without loss of generality that $\mu = c = 1$ in all proofs. 

The set of admissible policies $\mathbf{\Pi}$ is the set of non-anticipating policies. An admissible policy $\pi$ may be represented by a vector of arrival rates $\{\lambda_0, \lambda_1, \lambda_2, \dots \}$. From the memoryless property, it is sufficient to let the number of customers in the system represent the state of the system. When there are $i$ customers in the system, the price is set to $p(\lambda_i)$. In some instances, specifically when $\Lambda \geq 1$, some admissible policies induce unstable queues. To be precise, if we write $\mu_i = \min(i,C)$, and define $b_n = \Pi_{i=1}^{n} (\lambda_{i-1}/\mu_i)$, then the resulting queue is stable if and only if $\sum_{n=0}^\infty b_n$ converges (see e.g. pg. 254 of \cite{wolff1989stochastic}). Otherwise, the queue increases without bound. However, we can restrict ourselves, without loss of optimality, to considering policies which induce stable queues, as those which do not induce stability incur infinitely large costs. The stationary probabilities of a policy $\pi \in \mathbf{\Pi}$ are denoted by $\mathbb{P}_i(\pi)$ for each state  $i \in \{0,1,2\dots\}$. We denote the optimal policy by $\pi^*$ with corresponding optimal arrival rates of $\lambda_0^*,\lambda_1^*,\lambda_2^*,\ldots$ and steady-state probabilities $\mathbb{P}_0(\pi^*),\mathbb{P}_1(\pi^*),\mathbb{P}_2(\pi^*),\ldots$.    

The static policies we consider are those non-anticipating policies which fix a price up to a certain occupancy threshold (potentially infinite) and then disallow arrivals after that point. In other words, we consider the class of policies  that fix an arrival rate $\lambda$ in every state at or below a certain cutoff point $\gamma \in \{0,1,2,\dots\}$ and blocks arrivals in state $\gamma+1$ and higher by setting the arrival rates to $0$ (infinite price). An arbitrary static policy is notated $\pi^{\lambda, \gamma}$, which refers to the policy which picks the arrival rate $\lambda$ for customers whenever there are at most $\gamma$ customers in the system, and otherwise sets all arrival rates to $0$.

\subsection{Objectives}

For a given policy $\pi \in \mathbf{\Pi}$ which induces a stable queue, we define $\mathcal{R}(\pi)$ as the average expected revenue rate attained by $\pi$, i.e., 
\begin{align}\label{revenueDefinition}
\mathcal{R}(\pi) 
:= \sum_{i=0}^\infty \lambda_i p(\lambda_i)\mathbb{P}_i(\pi) .
\end{align}
For policies which do not induce stable queues, the stationary probabilities $\mathbb{P}_i(\pi)$ do not exist. However, the revenue rate of any policy is finite, bounded above by $\max_\lambda \lambda p(\lambda)$ which is the maximum achievable revenue rate.

Similarly, we define $\mathcal{C}(\pi)$ as the  ``congestion penalty'' incurred by policy $\pi$. When the congestion penalty is proportional to the number of customers in the system, we have
\begin{align*}
\mathcal{C}(\pi) 
:= \sum_{i=0}^\infty ci\mathbb{P}_i(\pi) = cE[L(\pi)]
\end{align*}
where $L(\pi)$ represents the stationary number-in-system when adhering to policy $\pi$. Similar to the revenue rate equation, this formula is not well-defined for policies which do not induce stability -  for a policy $\pi$ that does not satisfy the stability condition, then the queue increases without bound and  $\mathcal{C}(\pi) = \infty$. We consider the case when the congestion penalty is proportional to the expected sojourn time in Section \ref{sojournModel}.

Now we can write our overall objective function, $\mathcal{Z}(\pi)$, as
\begin{align*}
    \mathcal{Z}(\pi) := \mathcal{R}(\pi) - \mathcal{C}(\pi) = \sum_{i=0}^\infty \left( \lambda_i p(\lambda_i) - ci \right) \mathbb{P}_i(\pi) .
\end{align*}
The optimal policy $\pi^*$ can now be formalized as
\begin{equation}
\label{eq:optimal-policy}
    \pi^* \;:=\; \argmax_{\pi \in \Pi} \mathcal{Z}(\pi).
\end{equation}
\noindent
We denote $\lambda_i^*$ as the optimal arrival rate chosen by $\pi^*$ when there are $i$ customers 
in the system. An immediate question is whether the optimal arrival rates $\lambda_i^*$ 
satisfy a natural monotonicity in $i$. Proposition \ref{optimal-monotone} confirms that 
the optimal policy indeed prescribes lower arrival rates (higher prices) 
as the queue grows.

\begin{proposition}\label{optimal-monotone}
Let $\bar{\lambda} := \arg\max_{\lambda}\{\lambda\,p(\lambda)\}$ be the 
``myopic'' arrival rate. Then
\[
  \bar{\lambda} \;\ge\; \lambda_0^* \;\ge\; \lambda_1^* 
  \;\ge\; \lambda_2^* \;\ge\; \dots
\]
i.e., the optimal arrival rate is nonincreasing with respect to state $i$.
\end{proposition}
\proof{Proof.}
We begin formulating the problem as a continuous-time Markov decision process (MDP) 
via the uniformization technique. Specifically, let $h(i)$ denote the long-run expected 
reward when there are $i$ customers. From the definition of $\mathcal{Z}(\pi)$ above, the Bellman operator $\mathcal{T}$ is 
 
\begin{equation}
\label{MDP-equation}
\mathcal{T}h(i) := \max_{\lambda \in [0,\Lambda]} \Bigl\{ \lambda p(\lambda)-i  - \eta +
  \lambda \gamma h(i+1) + \gamma h(i-1) + [1 - \gamma(\lambda+1)] h(i)
\Bigr\},
\end{equation}
where $\gamma = \tfrac{1}{1 + \Lambda}$ dominates the total transition rate out of state~$i$ (to states $i-1,i,i+1$), and $\eta$ is the optimal average profit.

Let $h^*(i) := \lim_{n \to \infty} \mathcal{T}^n h(i)$. We will show that $h(\cdot)$ 
being nonincreasing and concave is preserved under $\mathcal{T}$, implying that $h^*(\cdot)$ is non-increasing and concave. From that, it follows that $\{\lambda_i^*\}$ (the arrival rates 
chosen by the optimal policy) must also be nonincreasing in $i$. 

We first decompose $\mathcal{T}h(i)$. Write $ \mathcal{T}h(i) \;=\; A(i) \;+\; B(i),$
where
\[
  A(i) \;=\; \max_{\lambda \in [0,\Lambda]}
  \Bigl\{\lambda p(\lambda)
   + \gamma\,\lambda\,h(i+1)
   + \gamma\,(\Lambda-\lambda)\,h(i)\Bigr\},
  \quad
  B(i) \;=\; \gamma\,h(i-1)\;- i - \;\eta.
\] 
One can immediately see that $B(\cdot)$ is nonincreasing and concave if $h(\cdot)$ is nonincreasing and concave. Let $\lambda_i := \arg\max A(i)$. $A(\cdot)$ is nonincreasing in since $h(\cdot)$ is non-increasing. To establish $A(\cdot)$ is convex, observe that 
\begin{align*}
  [A(i+1) - A(i)] - [A(i)-A(i-1)] 
  & \leq A(i-1)|_{\lambda_{i-1}} + A(i+1)|_{\lambda_{i+1}} - A(i)|_{\lambda_{i-1}} - A(i)|_{\lambda_{i+1}} \\
    &=   \gamma  (\Lambda-\lambda_{i-1}) (h(i+1)+h(i-1)-2h(i))  \\
    &+ \gamma \lambda_{i+1} (h(i+2)+h(i)-2h(i+1))\\
    &\leq 0.
\end{align*}
The first inequality follows from optimality of $\lambda_i$ for $A(i)$ and the second inequality follows from the  concavity of $h(\cdot)$.

Finally, the optimal arrival rates satisfy
\[
  \lambda_i^* 
  \;=\;
  \arg\max_{\lambda}\,
  \Bigl\{ \lambda \left[ p(\lambda) + \gamma (h^*(i+1) - h^*(i-1)) \right]
  \Bigr\}.
\]
Because $h^*(\cdot)$ is nonincreasing and concave, the differences $h^*(i+1)-h^*(i)$ ensures 
$\lambda_i^*$ forms a nonincreasing sequence with $i$. In addition, $\lambda_0^*$ 
cannot exceed the myopic best response $\bar{\lambda} = \arg\max_{\lambda}\{\lambda\,p(\lambda)\}$. 
Hence, $  \bar{\lambda} \;\ge\; \lambda_0^*  \;\ge\; \lambda_1^*  \;\ge\; \dots$ as stated in the claim.
\Halmos
\endproof

\color{black}
\subsection{Little's Law and Static Policies}
The well-known Little's Law is integral to our work and helps motivate the static policy construction we use to prove our guarantees. Suppose we fix an instance of our model, and consider an optimal policy $\pi^*$. The policy $\pi^*$ induces a distribution $L(\pi^*)$ of the number of customers in the system in stationarity. Moreover, when a customer joins the system in stationarity, they will experience a sojourn time distributed according to a random variable we denote by $W(\pi^*)$. Little's Law allows us to relate the first moments of these two distributions, i.e., 
\begin{align*}
    E[L(\pi^*)] = \tilde\lambda E[W(\pi^*)]
\end{align*}
where $\tilde\lambda$ denotes the long-run average arrival rate of customers  under $\pi^*$. 

We can compute this average arrival rate of customers under the optimal policy $\pi^*$ using the stationary probabilities, i.e., 
\begin{align*}
\tilde\lambda = \sum_{i = 0}^\infty \lambda_i^* \mathbb{P}_i(\pi^*).
\end{align*}
Our key proof idea is to use this average arrival rate to construct good static policies. In any state in which our static policy sells, we pick the arrival rate $\tilde\lambda$. We note that this differs from the static policy used in \cite{doi:10.1287/opre.2020.2054} to attain performance guarantees for the Erlang loss model. Their static policy  was also derived from the optimal policy although it was based on a modification of the expected arrival rate, and the analysis is quite different than ours due to the mixed objective function $\mathcal{Z}(\pi)$ we study.

To give some intuition behind our construction, the idea behind picking $\tilde\lambda$ as the arrival rate used by our static policies is that we are trying to construct a static policy which behaves similarly to the optimal dynamic policy. If the optimal dynamic policy has a low average arrival rate, it suggests that customers worth admitting are few and far between, and so a good static policy should likely also pick a low rate. On the other hand, if the average arrival rate of the optimal policy is high, then it stands to reason that a good static policy will also seek to harvest revenue from many of the arriving customers, and thus will select a higher arrival rate. The fact that we can prove guarantees for our static policies which pick $\tilde\lambda$ lends credence to this intuition.
\section{Single-server Static Pricing Guarantees}\label{StaticCutoffSection}
We begin with the single-server case, which has simple, closed-form results and allows us to  illustrate our ideas before moving to multiple servers. We next present our results and defer the proofs and tight examples to later subsections. Our first result in Theorem \ref{singleServerProfit} is a 50\% universal guarantee on the overall objective rate of the policy $\pi^{\tilde\lambda, 0}$, which sets the arrival rate to be $\tilde\lambda = \sum_{i = 0}^\infty \lambda_i^* \mathbb{P}_i(\pi^*)$ and only allows customers to join the queue when it is empty. 

\begin{theorem}\label{singleServerProfit}
    Let $C=1$ and let $\pi^*$ be an optimal dynamic policy which induces an average arrival rate $\tilde\lambda$. For any $\Lambda>0$, $\mu>0$, and regular $F$,  the policy $\pi^{\tilde\lambda,0}$ attains at least half of the optimal objective value. Equivalently,
    \begin{align*}
    \mathcal{Z}(\pi^{\tilde\lambda,0}) \geq \frac{\mathcal{Z}(\pi^*)}{2}.
    \end{align*}
    Moreover, there is a problem instance showing that our analysis is tight.
\end{theorem}

As Theorem \ref{singleServerProfit} asserts, there are instances where the $\frac{1}{2}$ guarantee is the best one can do with policies using the static arrival rate $\tilde\lambda$. 
Here, however, it is important to note that we introduce $\tilde\lambda$---the average arrival rate under the optimal dynamic policy---\emph{only} as a proof construct. In other words, we do \emph{not} advocate that practitioners determine $\tilde\lambda$ in order to implement $\pi^{\tilde\lambda,0}$. Instead, $\tilde\lambda$ is a theoretical device used to prove that such a static policy must exist. In practice, one would simply choose a price (i.e., an arrival rate) and cutoff via a straightforward two-parameter search, rather than relying on any knowledge of $\tilde\lambda$ from the dynamic model.

Moreover, in Section \ref{tightnessExample}, we provide instances such that $\mathcal{Z}(\pi^{\tilde\lambda, \gamma}) < 0$ for any $\gamma > 0$ despite having $\mathcal{Z}(\pi^*) > 0$, which  means we cannot provide objective value guarantees for arbitrary cutoffs. In light of this, we consider what happens when we allow policies to let queues form. Theorem \ref{StaticGuarantee}  gives tight bounds on the approximation ratio for revenue and queue length simultaneously. Moreover, we now allow the cutoff threshold to vary and get a family of guarantees.

\begin{theorem}\label{StaticGuarantee}
Let $C=1$ and let $\pi^*$ be an optimal dynamic policy which induces an average arrival rate $\tilde\lambda$ and consider any $\Lambda>0$, $\mu>0$, and regular $F$. Then for any $\gamma \in \{0,1,2,3,\dots\}$, the static pricing policy $\pi^{\tilde\lambda, \gamma}$ guarantees at least $\frac{\gamma+1}{\gamma+2}$ of the revenue rate and incurs a queue length (cost) at most $g(\gamma)$ times that of the optimal dynamic pricing policy, where 
\[
  g(\gamma) = \begin{cases}
      1, & \text{if } \gamma=0\\
      \frac{2}{\sqrt{3}}, & \text{if } \gamma =1 \\
      1.532, & \text{if } \gamma = 2\\
      \frac{\gamma+1}{2}, & \text{if } \gamma \geq 3
  \end{cases}
\]
Equivalently, we simultaneously guarantee that
\begin{align*}
    \frac{\gamma+1}{\gamma+2}\mathcal{R}(\pi^*)\;\leq\;\mathcal{R}(\pi^{\tilde\lambda, \gamma})
    \quad\text{and}\quad
    g(\gamma)\mathcal{C}(\pi^*)\;\geq\;\mathcal{C}(\pi^{\tilde\lambda, \gamma}).
\end{align*}
Moreover, for each $\gamma$, there is a problem instance showing that our analysis is tight. 
\end{theorem}

Theorem \ref{StaticGuarantee} provides a family of guarantees, parameterized by the value of $\gamma$. This allows the practitioner to select a value of $\gamma$ suitable to their application: for scenarios in which revenues are expected to be very large compared to costs, a higher value of $\gamma$ is warranted, and \emph{vice versa} when costs are expected to be large relative to revenues. Moreover, we will see that the same class of instances which demonstrates the tightness of Theorem \ref{singleServerProfit} proves that Theorem \ref{StaticGuarantee}'s guarantees are tight as well.

Finally, it is important to reiterate that while Theorems~\ref{singleServerProfit} and~\ref{StaticGuarantee} utilize $\tilde{\lambda}$, we do not propose that one \emph{compute} this value in order to implement a static policy. Our constructions serve only to show that suitable two-parameter policies (involving a single price/arrival rate and a cutoff) \emph{must} exist and achieve these performance bounds. In an actual implementation, the search over these two parameters is straightforward and does not require the infinite-dimensional optimization of the dynamic problem. Hence, our proofs are constructive---they show existence---but they are not intended as a prescriptive algorithm to find static policies by first determining $\tilde{\lambda}$.

The fact that we can provide exact\footnotemark \footnotetext{The exact, closed-form expression for $g(2)$ involves the root of a quartic and would take a few pages to relay.} values for the function $g(\gamma)$ is a stroke of some good luck. As will be seen in the proof, finding these worst-case cost bounds involves solving an optimization problem over $\tilde\lambda \in [0,1]$. For $0 < \gamma < 3$, the maximizer lies strictly within the interval $[0,1]$; for $\gamma \geq 3$, the function is monotonically increasing on the interval and hence the maximum occurs when $\tilde\lambda = 1$. This is fortunate because finding the \textit{unconstrained} maxima of the relevant ratio involves finding the roots higher and higher degree polynomials: for $\gamma = 1$, a degree $2$ polynomial arises, for $\gamma = 2$, a degree $4$ polynomial arises, and for general $\gamma$ a degree $2\gamma$ polynomial arises. If the unconstrained maximum when $\gamma=3$ were to lie in the interval $[0,1]$, it is unlikely we could find a closed-form expression.

We first prove the profit rate result in Section \ref{ProfitRate}. In Sections \ref{RevRate} and \ref{CostRate}, we establish the revenue and cost rate guarantees, respectively. Later, in Section \ref{tightnessExample}, we describe the class of instances which will show our guarantees are tight.

For an arbitrary policy $\pi$, we can compute the steady state probabilities $\mathbb{P}_i(\pi)$ as
\begin{equation}\label{steadyStateProbabilities}
    \mathbb{P}_i(\pi) = \frac{\prod_{k=0}^{i-1} \lambda_k}{1+\sum_{n=0}^\infty \prod_{k=0}^n \lambda_k }, \quad i=0,1,2,3,\dots
\end{equation}
where $\prod_{k=0}^{-1}\lambda_k = 1$. For policies which induce a null recurrent or transient process, these probabilities are not defined. In fact, by applying Theorem 2 of \cite{Karlin1957TheCO}, we get that the induced process is ergodic precisely when the denominator of Equation (\ref{steadyStateProbabilities}) is finite. Since any policies which do not induce ergodic processes incur arbitrarily large costs, we assume our policies are such that $\sum_{n=0}^\infty\prod_{k=0}^n\lambda_k < \infty$.

We make use of the following bound on the average arrival rate induced by the optimal policy.

\begin{lemma}\label{singleserverstability} For any policy $\pi = \{\lambda_0, \lambda_1, \dots \}$ which induces a stable queue,
$\tilde{\lambda} < 1$.
\end{lemma}
\proof{Proof.}
We can write
\begin{align*}
\tilde{\lambda} = \sum_{i=0}^\infty \lambda_i \mathbb{P}_i(\pi)= \sum_{i=0}^\infty \lambda_i\frac{\prod_{k=0}^{i-1} \lambda_k}{1+\sum_{n=0}^\infty \prod_{k=0}^n \lambda_k }= \frac{\sum_{i=0}^\infty\prod_{k=0}^{i} \lambda_k}{1+\sum_{n=0}^\infty \prod_{k=0}^n \lambda_k }< 1
\end{align*} 
where the first equality is by definition, the second equality is from \eqref{steadyStateProbabilities}, and the inequality uses the fact that the denominator is finite by the stability assumption. 
\Halmos \endproof

We make one more useful observation. Consider the static policy $\pi^{\lambda, \gamma}$ which sets an arrival rate $\lambda$ for all states below and including $\gamma$, and cuts off arrivals by setting all others to $0$. When we fix such a policy, the queueing part of the model becomes a classical $M/M/1/\gamma+1$ model with service rate $1$ and arrival rate $\lambda$. Thus we get the following well-known form of the stationary probabilities (see e.g. \cite{1990665}) which will be used in our arguments

\begin{align}\label{SSprobMM1C+1}
\mathbb{P}_i(\pi^{\lambda,\gamma}) = \begin{cases}
\Big(\frac{1-\lambda}{1-\lambda^{\gamma+2}}\Big)\lambda^i  & (\lambda \neq 1) \\
\frac{1}{\gamma+2} & (\lambda=1)
\end{cases}
\end{align}
\subsection{Approximation of profit rate}\label{ProfitRate}
We now show that $\pi^{\tilde\lambda,0}$ attains at least half of the optimal profit,  proving Theorem \ref{singleServerProfit}.
\proof{Proof of Theorem \ref{singleServerProfit}.}
    Using Eq. (\ref{revenueDefinition}), we first arrive at an upper bound for the revenue rate of the optimal policy
\begin{equation}\label{RevBound}
    \mathcal{R}(\pi^{*}) = \sum_{i=0}^\infty \lambda_i^*p(\lambda_i^*)\mathbb{P}_i(\pi^*)\leq \tilde{\lambda}p(\tilde{\lambda})
\end{equation}
The inequality follows by leveraging the concavity of the revenue rate function to apply Jensen's inequality to a random variable $X$ with probability mass function $P(X=\lambda_i^*)=\mathbb{P}_i(\pi^*)$. 

Now we compute the revenue rate of the policy $\pi^{\tilde\lambda,0}$ using Eqs. (\ref{revenueDefinition}) and (\ref{SSprobMM1C+1}) to arrive at a lower bound for $\mathcal{R}(\pi^{\tilde\lambda,0}) $
\begin{align}\label{revBoundForProfit}
\mathcal{R}(\pi^{\tilde\lambda,0}) &= \tilde\lambda p(\tilde\lambda) \mathbb{P}_0(\pi^{\tilde\lambda,0}) \nonumber\\
&= \tilde\lambda p(\tilde\lambda) \frac{1}{1+\tilde\lambda}\nonumber\\
&\geq \mathcal{R}(\pi^*) \frac{1}{1+\tilde\lambda}
\end{align}
where the inequality follows from Eq. (\ref{RevBound}).

We now turn to cost rates. We first apply Little's Law to lower bound the optimal cost rate $\mathcal{C}(\pi^*)$. Letting $L^*$ be distributed according to the optimal stationary distribution given by the $\mathbb{P}_i^*$, we have
\begin{align}\label{LittlesLaw1}
\mathcal{C}(\pi^*) = E[L(\pi^*)] = \tilde\lambda E[W(\pi^*)] \geq \tilde\lambda
\end{align}
where $W^*$ denotes the stationary sojourn time distribution of a customer under $\pi^*$. The inequality follows from observing that $E[W(\pi^*)] \geq 1$ because any sojourn includes a service time.

We now seek an upper bound on the cost rate of the policy $\pi^{\tilde\lambda,0}$. The expected cost rate is simply the proportion of time a customer is in the system, which we compute with Eq. (\ref{SSprobMM1C+1})
\begin{align*}
\mathcal{C}(\pi^{\tilde\lambda,0}) = \mathbb{P}_1(\pi^{\tilde\lambda,0})=\frac{\tilde\lambda}{1+\tilde\lambda}
\end{align*}
Combining this with Eq. (\ref{LittlesLaw1}) we get that
\begin{align}\label{costBoundForProfit}
\mathcal{C}(\pi^{\tilde\lambda,0}) \leq \mathcal{C}(\pi^*)\frac{1}{1+\tilde\lambda}.
\end{align}
We can now use the fact that $\mathcal{P}(\pi^*) = \mathcal{R}(\pi^*)-\mathcal{C}(\pi^*)$ combined with Eqs. (\ref{revBoundForProfit}) and (\ref{costBoundForProfit}) to lower bound the profit rate of $\pi^{\tilde\lambda,0}$:
\begin{align*}
\mathcal{Z}(\pi^{\tilde\lambda,0}) = \mathcal{R}(\pi^{\tilde\lambda,0}) - \mathcal{C}(\pi^{\tilde\lambda,0}) \geq \frac{\mathcal{R}(\pi^*)}{1+\tilde\lambda} - \frac{\mathcal{C}(\pi^*)}{1+\tilde\lambda} = \frac{\mathcal{Z}(\pi^*)}{1+\tilde\lambda}.
\end{align*}
Since $\tilde\lambda < 1$ by Lemma \ref{singleserverstability}, we have the result
\begin{align*}
\mathcal{Z}(\pi^{\tilde\lambda, 0})\geq \frac{\mathcal{Z}(\pi^*)}{2} . \Halmos
\end{align*}
 \endproof

\subsection{Approximation of revenue rate}\label{RevRate}
We now state and prove Lemma \ref{RevRateC=0Prop} which ensures our constructed policy attains the revenue guarantee in Theorem \ref{StaticGuarantee}.
\begin{lemma}\label{RevRateC=0Prop}Suppose $\pi^*$ is an optimal dynamic policy and that it induces the average arrival rate $\tilde\lambda$. Then, for any $\gamma$, the policy $\pi^{\tilde\lambda, \gamma}$ satisfies
\begin{align*}
    \frac{\gamma+1}{\gamma+2}\mathcal{R}(\pi^*)\leq \mathcal{R}(\pi^{\tilde\lambda, \gamma})
\end{align*}
\end{lemma}
\proof{Proof.}
    Using Eq. (\ref{revenueDefinition}), we can compute the revenue rate of the static policy $\pi^{\tilde\lambda,\gamma}$:
\begin{align}\label{RevBoundInitial}
    \mathcal{R}(\pi^{\tilde{\lambda},\gamma}) &=\sum_{i=0}^\infty \tilde{\lambda} p(\tilde\lambda) \mathbb{P}_i(\pi^{\tilde{\lambda},\gamma}) =  \tilde{\lambda}p(\tilde{\lambda})(1-\mathbb{P}_{\gamma+1}(\pi^{\tilde{\lambda},\gamma}))
\end{align}

Using the facts that $\tilde{\lambda} < 1$ and that the blocking probability $\mathbb{P}_{\gamma+1}(\pi^{\lambda,\gamma})$ is increasing in $\lambda$, we get the following bound:
\begin{align*}
    \mathbb{P}_{\gamma+1}(\pi^{\tilde{\lambda},i})\leq \mathbb{P}_{\gamma+1}(\pi^{1,i}) = \frac{1}{\gamma+2}
\end{align*}

Combining this inequality with Eqs. (\ref{RevBound}) and (\ref{RevBoundInitial}) furnishes the desired revenue guarantee:
\begin{align*}
\mathcal{R}(\pi^{\tilde{\lambda},\gamma})=\tilde{\lambda}p(\tilde{\lambda})(1-\mathbb{P}_{\gamma+1}(\pi^{\tilde{\lambda},\gamma})) \geq \mathcal{R}(\pi^*) \frac{\gamma+1}{\gamma+2} \Halmos
\end{align*} \endproof

\subsection{Approximation of cost rate}\label{CostRate}



We now turn to the question of cost and prove the cost guarantees of Theorem \ref{StaticGuarantee}.  
\begin{lemma}\label{CostRateSingleServer}
Suppose $\pi^*$ is an optimal dynamic policy and that it induces the average arrival rate $\tilde\lambda$. Then for any $\gamma$, the policy $\pi^{\tilde\lambda, \gamma}$ satisfies
\begin{align*}
    g(\gamma) \mathcal{C}(\pi^*)\geq \mathcal{C}(\pi^{\tilde\lambda, \gamma})
\end{align*}
where 
  \[
    g(\gamma) = \begin{cases}
        1, & \text{if } \gamma=0\\
        \frac{2}{\sqrt{3}}, & \text{if } \gamma =1 \\
        1.532, & \text{if } \gamma = 2\\
        \frac{\gamma+1}{2}, & \text{if } \gamma \geq 3
        \end{cases}.
  \]
\end{lemma}

\proof{Proof.}
We compute the cost rate of the policy $\pi^{\tilde\lambda, \gamma}$ and arrive at an upper bound in terms of the optimal cost 
\begin{align}
\mathcal{C}(\pi^{\tilde\lambda, \gamma}) &= \sum_{i=1}^{\gamma+1} i \mathbb{P}_i(\pi^{\tilde\lambda, \gamma}) \nonumber \\
&= \frac{\mathcal{C}(\pi^*)}{\mathcal{C}(\pi^*)}\sum_{i=1}^{\gamma+1} i \mathbb{P}_i(\pi^{\tilde\lambda, \gamma}) \nonumber \\
&\leq \mathcal{C}(\pi^*) \frac{\sum_{i=1}^{\gamma+1} i \mathbb{P}_i(\pi^{\tilde\lambda, \gamma}) }{\tilde\lambda} \label{theOneInequality}
\end{align}
where the final inequality uses Eq. (\ref{LittlesLaw1}).

From Lemma \ref{singleserverstability}, we know that $\tilde{\lambda} <1$. Thus, for fixed $\gamma$ we can now furnish the worst-case cost bound by finding the maximum value of the coefficient of $\mathcal{C}(\pi^*)$ for $\tilde\lambda \in [0,1)$. Specifically, we will let \begin{align*} g(\gamma) = \max_{\tilde\lambda \in [0,1]} \frac{\sum_{i=1}^{\gamma+1} i \mathbb{P}_i(\pi^{\tilde\lambda, \gamma}) }{\tilde\lambda} 
\end{align*}
in which case we will have 
\begin{align*}
\mathcal{C}(\pi^{\tilde\lambda, \gamma}) \leq \mathcal{C}(\pi^*) g(\gamma)
\end{align*}
which is the desired inequality.
For $\gamma < 3$ it is straightforward to arrive at the claimed values of $g(\gamma)$. For example, when $\gamma=2$, finding the maximizer reduces to finding the real positive root of the equation
\begin{align*}
-3\tilde\lambda^4-4\tilde\lambda^3-2\tilde\lambda^2+4\tilde\lambda+1
\end{align*}
which can be accomplished via the well-known Quartic formula due to Ferrari \cite{ArsMagna}.

For $\gamma \geq 3$, we proceed by showing that the objective is increasing in $\tilde\lambda$ over the interval $[0,1]$. Let $f_\gamma(\tilde\lambda)$ denote the objective function for a fixed $\gamma$ and express it equivalently as
\begin{align*}
f_\gamma(\tilde\lambda) = \frac{\sum_{j=1}^{\gamma+1}j\tilde\lambda^j}{\sum_{i=1}^{\gamma+2}\tilde\lambda^i}
\end{align*}
We  show the derivative $f'_\gamma(\tilde\lambda)$ is nonnegative for $0\leq\tilde\lambda\leq1$ and $\gamma \geq 3$. We take the derivative and ignore the denominator by looking only at the signs
\begin{align*}
\text{sgn}(f'_\gamma(\tilde\lambda)) &= \text{sgn}\left(\sum_{i=1}^{\gamma+2}\sum_{j=0}^{\gamma+1}j^2\tilde\lambda^{i+j-1} - \sum_{i=1}^{\gamma+2}\sum_{j=0}^{\gamma+1}ji\tilde\lambda^{i+j-1}\right) \\
&= \text{sgn}\left(\sum_{i=0}^{\gamma+1}\sum_{j=0}^{\gamma+1}j^2\tilde\lambda^{i+j} - \sum_{i=0}^{\gamma+1}\sum_{j=0}^{\gamma+1}j(i+1)\tilde\lambda^{i+j}\right) \\
&= \text{sgn}\left(\sum_{i=0}^{\gamma+1}\sum_{j=0}^{\gamma+1}(j^2 -j(i+1))\tilde\lambda^{i+j}\right) 
\end{align*}
We  show this is nonnegative for $0 \leq \tilde\lambda \leq 1 $ when $\gamma\geq3$ by showing that the coefficients of the lower-order terms (those with degrees lesser or equal to $\gamma+1$) are all positive, and that the sum of all coefficients is positive as well. This suffices because for $\tilde\lambda\leq1$, the lower-order terms dominate.

We can use the above expression to write the coefficient of $\tilde\lambda^k$ for $2\leq k\leq\gamma+1$ as 
\begin{align*}
\sum_{j=0}^k (j^2-j(k-j+1)) = \frac{(k-1)k(k+1)}{6}
\end{align*}
which we observe is indeed positive for $k\geq 2$. Proceeding similarly for the higher-order terms, for $\gamma+1<k\leq2\gamma+2$ the coefficient of $\tilde\lambda^k$ is
\begin{align*}
\sum_{j=k-\gamma+1}^{\gamma+1} (j^2-j(k-j+1)) = \frac{(2\gamma-k+3)(4\gamma^2-4\gamma k+12\gamma +k^2-9k+8)}{6}.
\end{align*}
To conclude, we show that the sum of all the coefficients is nonnegative for $\gamma\geq 3$. Indeed,

\begin{align*}
\sum_{k=2}^{\gamma+1} \frac{(k-1)k(k+1)}{6} 
&+ \sum_{k=\gamma+2}^{2\gamma+2} 
   \frac{(2\gamma-k+3)(4\gamma^2 - 4\gamma k + 12\gamma + k^2 - 9k + 8)}{6} \\
&= \frac{1}{12} (\gamma+1)(\gamma+2)^2(\gamma-3).
\end{align*}

from which we can immediately see the desired nonnegativity. With this monotonicity in hand, for $\gamma \geq 3$,  we can simply evaluate the objective function at $\tilde\lambda = 1$  
\begin{align*}
g(\gamma) = \frac{\sum_{i=1}^{\gamma+1} i \mathbb{P}_i(\pi^{1, \gamma}) }{1} = \sum_{i=1}^{\gamma+1} \frac{i}{\gamma+2} = \frac{\gamma+1}{2},
\end{align*}
thus completing the proof.
\Halmos \endproof

\subsection{A class of instances proving tightness}\label{tightnessExample}
We now describe a class of instances which we use to show that Theorem \ref{singleServerProfit} and Theorem \ref{StaticGuarantee} are tight. This class of instances will also prove that no guarantee on the objective value approximation ratio like Theorem \ref{singleServerProfit} is possible when $\gamma>0$. Moreover, the following examples also show that static policies with no threshhold can perform arbitrarily poorly.

Consider an instance with linear demand (corresponding to a uniform valuation distribution which is regular), i.e. where $\lambda (p) = \max \{b-ap,0 \}$ for some positive $a$ and $b$. Take $c=\mu=1$ and take $a < b < 2a$, so that the expected value from accepting a customer when one is in service is always negative: the largest price a customer will pay is $b/a<2$, while the expected congestion penalty from accepting a customer while one is in service is $2$ (2 customers). Thus, the optimal policy is in fact a static policy with a cutoff point of $0$. In this case, it is a straightforward single-variable optimization problem and we can arrive at the following expression for the optimal policy $$\pi^* = \{\sqrt{b-a+1}-1,0,0,\dots \}$$ 

To now show the tightness of the objective value guarantee of Theorem \ref{singleServerProfit}, we write $b = \kappa a$ for $1 < \kappa < 2$ and take the limit as $a \rightarrow \infty$. In the limit, $\sqrt{\kappa a-a+1}-1 \rightarrow \infty$, and thus $\tilde\lambda \rightarrow 1$. Then we have 
\begin{align*}\mathcal{Z}(\pi^{\tilde\lambda, 0})  = \tilde\lambda p(\tilde\lambda) \mathbb{P}_0(\pi^{\tilde\lambda, 0}) - \mathbb{P}_1(\pi^{\tilde\lambda, 0})  =\tilde\lambda p(\tilde\lambda) \frac{1}{1+\tilde\lambda} - \frac{\tilde\lambda}{1+\tilde\lambda} = \tilde\lambda\frac{(\kappa a - \tilde\lambda)}{a(1+\tilde\lambda)} - \frac{\tilde\lambda}{1+\tilde\lambda} \rightarrow \frac{\kappa-1}{2} \end{align*}
On the other hand, we can see that the optimal price converges to $\kappa$
$$p(\lambda_0^*) = \frac{\kappa a-\sqrt{\kappa a-a+1}+1}{a} \rightarrow \kappa$$
and thus
\begin{align*}\mathcal{Z}(\pi^*) &= \lambda_0^* p(\lambda_0^*) \mathbb{P}_0(\pi^*)  - \mathbb{P}_1(\pi^*)  = \frac{\lambda_0^*}{1+\lambda_0^*}p(\lambda_0^*) - \frac{\lambda_0^*}{1+\lambda_0^*} \rightarrow \kappa-1 \end{align*}
which proves the desired tightness for $\gamma=0$.
Moreover, since the expected value of allowing a customer to queue is always negative, policies $\pi^{\tilde\lambda, \gamma}$ with $\gamma > 0$ will perform worse, and thus we observe that in this instance the best static policy which fixes an arrival rate $\tilde\lambda$ attains only half the optimal value. However, it is worth noting that in this instance the optimal policy is in fact a static policy, just one which does not pick the arrival rate $\tilde\lambda$.

The same limiting instance shows the revenue guarantees of Theorem \ref{StaticGuarantee} are also tight. For any $\gamma$, we have
\begin{align*}
\mathcal{R}(\pi^{\tilde\lambda, \gamma}) = \tilde\lambda p(\tilde\lambda)(1-\mathbb{P}_{\gamma+1}(\pi^{\tilde\lambda, \gamma})) \rightarrow \kappa\frac{\gamma+1}{\gamma+2}
\end{align*}
On the other hand, as we computed above,
\begin{align*}
\mathcal{R}(\pi^*) = \lambda_0^* p(\lambda_0^*) \mathbb{P}_0(\pi^*) \rightarrow \kappa
\end{align*}
which proves the claimed tightness of the revenue guarantees.

The same class of instances will also serve to prove that the congestion penalty guarantees of Theorem \ref{StaticGuarantee} are tight as well, but we will not use the limiting instance. Instead, note that the only inequality used to furnish our congestion penalty guarantees is Eq. (\ref{theOneInequality}). Since for instances in this class the optimal policy $\pi^*$ does not allow a queue to form, this inequality is tight, and hence so are our ensuing guarantees. 

These instances are also such that $\mathcal{Z}(\pi^{\tilde\lambda, \gamma}) < 0$ for any $\gamma > 0$, which is why we cannot provide objective value guarantees for arbitrary cutoffs and must resort to giving the separate, bi-criteria bounds of Theorem \ref{StaticGuarantee}. Indeed, we can compute
\begin{align*}
\mathcal{Z}(\pi^{\tilde\lambda, \gamma}) &= \mathcal{R}(\pi^{\tilde\lambda, \gamma}) - \mathcal{C}(\pi^{\tilde\lambda, \gamma})\\
&\rightarrow \frac{\gamma+1}{\gamma+2}\kappa - \frac{\gamma+1}{2} \end{align*}
where the limit is again taken as $a \rightarrow \infty$. Now note that this value is negative whenever $\kappa < (\gamma+2)/2$, and we can find such a $\kappa \in (1,2)$ precisely when $\gamma > 0$.

Finally, these same instances can be used to show that unthresholded static policies can perform arbitrarily poorly. Note that computing the optimal unthresholded policy for a given instance is a simple single-variable optimization problem. By taking $a$ large and $\kappa$ close to $1$, we can exhibit instances where the unthresholded static policy recovers only an $\epsilon$ proportion of the optimal value, for $\epsilon$ arbitrarily small. For example, with $a=1000$ and $\kappa = 1.05$, the optimal policy earns $0.37$ while the optimal unthresholded policy earns $0.0006$, which gives a performance ratio of $\frac{0.0006}{0.37} = 1.6\%$.

\section{Multi-server Static Pricing Guarantees}\label{multiServerSection}
Now we extend our results to the case when we have $C$ servers, each operating at rate $\mu$. All other aspects of the model remain the same. We now generalize Theorem \ref{singleServerProfit} to the multi-server case, again taking $\tilde\lambda = \sum_{i=0}^\infty \lambda_i^* \mathbb{P}_i^*$.
\begin{theorem}\label{multiServerProfit}
    For any $\Lambda >0, \mu>0$, and regular $F$,  the policy $\pi^{\tilde\lambda, C-1}$ achieves at least $1-\frac{\frac{1}{C!}(C^C)}{\sum_{n=0}^C\frac{1}{n!}C^n}$ of the profit of the optimal dynamic policy $\pi^*$. Equivalently,
    \begin{align*}
    \mathcal{Z}(\pi^{\tilde\lambda,C-1}) \geq \mathcal{Z}(\pi^*) \Bigg(1-\frac{\frac{1}{C!}(C^C)}{\sum_{n=0}^C\frac{1}{n!}C^n}\Bigg)
    \end{align*}
\end{theorem}

Theorem \ref{multiServerProfit} sheds light on how the performance of static pricing changes as we increase the number of servers. We observe the guarantees are strictly improving: for example, if we have ten servers, the theorem ensures the existence of a static policy which attains at least 78\% of the optimal profit (instead of the 50\% guarantee in the single-server case). With $100$ servers, the static policy $\pi^{\tilde\lambda,99}$ is guaranteed to achieve at least $92.4\%$ of the optimal profit. Taking this further, though the focus of the current work is on exact, non-asymptotic analysis, it is worth observing that the profit guarantee of Theorem \ref{multiServerProfit} tends to $1$ in the limit as $C \rightarrow \infty$. For systems with many servers, the advantage of dynamic pricing vanishes, and the performance of even our particular static policy becomes optimal.

To understand how our policies perform in the multi-server case when a queue is allowed to form, we provide the following bi-criteria approximations which generalize those in Theorem \ref{StaticGuarantee}.
\begin{theorem}\label{StaticGuaranteeMultiserverWithK}

Consider any $\Lambda >0, \mu>0$, and regular $F$. For any  $\gamma \in \{C-1,C,C+1,\ldots\}$, the static pricing policy $\pi^{\tilde\lambda,\gamma}$ guarantees at least $\Bigg(1-\frac{\frac{C^{C}}{C!}}{\sum_{l=0}^C\frac{1}{l!}C^l + \frac{1}{C!}C^C (\gamma+1-C)}\Bigg)$ of the revenue rate and incurs a cost rate less than $g(\gamma,C)$ times that of the optimal dynamic pricing policy, where 
\begin{align*}
g(\gamma, C) = \max_{\tilde\lambda \in [0,C]}\frac{\sum_{i=1}^{\gamma+1}i\mathbb{P}_i(\pi^{\tilde\lambda, \gamma})}{\tilde\lambda}.
\end{align*} Equivalently, 
\begin{align*}
    \Bigg(1-\frac{\frac{C^{C}}{C!}}{\sum_{l=0}^C\frac{1}{l!}C^l + \frac{1}{C!}C^C (\gamma+1-C)}\Bigg)\mathcal{R}(\pi^*)\leq \mathcal{R}(\pi^{\tilde\lambda, \gamma})\\
    g(\gamma,C) \mathcal{C}(\pi^*)\geq  \mathcal{C}(\pi^{\tilde\lambda, \gamma}).
\end{align*}
\end{theorem}

\begin{figure}[h!]
\includegraphics[scale=.8]{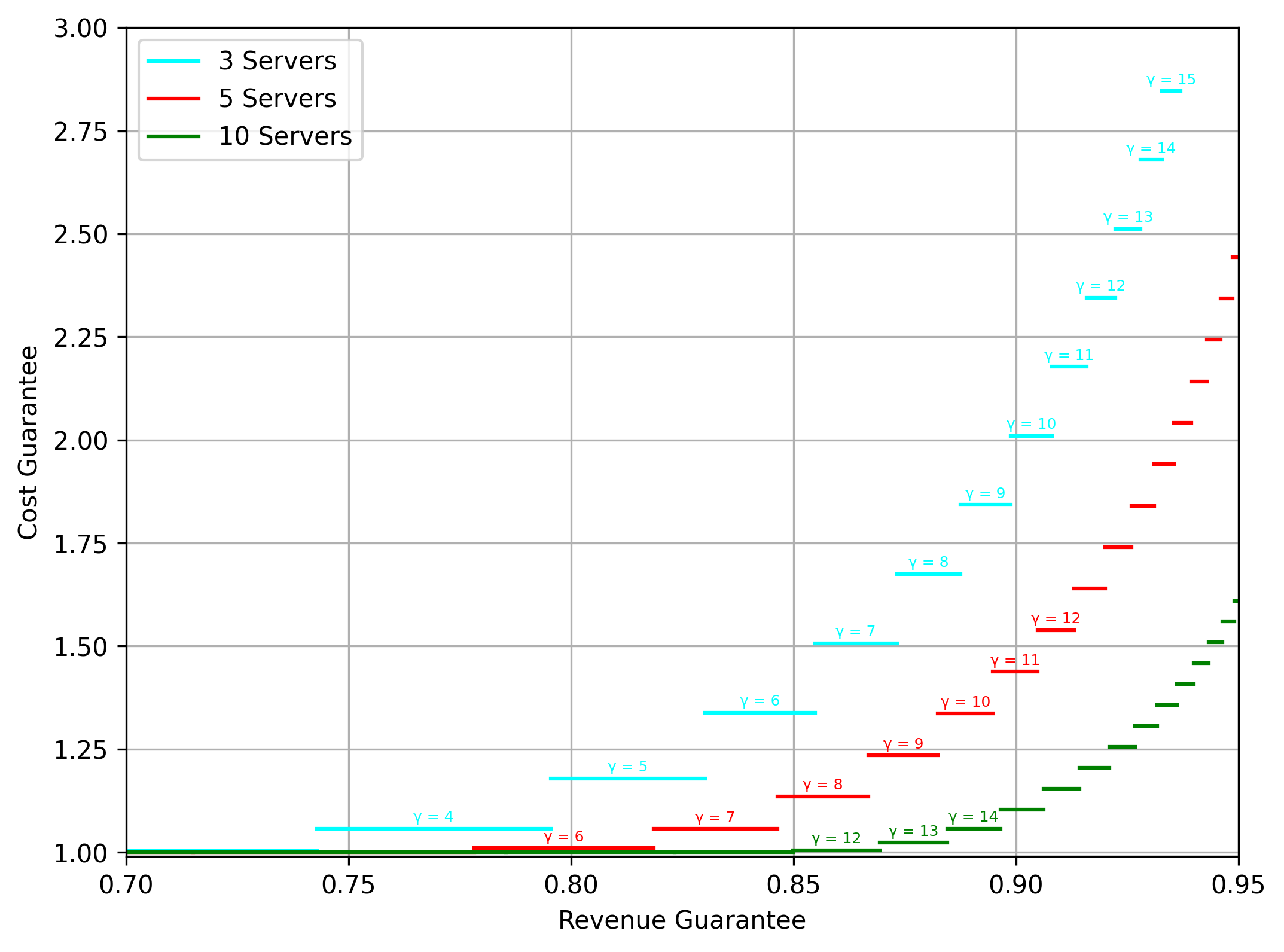} 
\centering
\caption{A graphical representation of the guarantees of Theorem \ref{StaticGuaranteeMultiserverWithK}}\label{fig1}
\end{figure}

When the number of servers increases,   the advantage of dynamic pricing over static pricing diminishes. Theorem \ref{StaticGuaranteeMultiserverWithK}  gives insight into the rate at which the advantage declines. Figure \ref{fig1} graphically shows the guarantees of Theorem \ref{StaticGuaranteeMultiserverWithK} when we have $3, 5,$ or $10$ servers. We recommend interpreting it as follows: first, fix a number of servers such as $C=3$. Now, determine a desired revenue guarantee. For example, suppose we want to ensure our static policy earns at least $80\%$ of the optimal revenue. Then following the line $x=0.80$ up until it intersects the cyan line, we can see that  by taking $\gamma=5$,  Theorem \ref{StaticGuaranteeMultiserverWithK} furnishes a static policy attaining at least $80\%$ of the optimal policy's revenue while incurring at most $1.18$ times the optimal cost (more precisely, moving to the right end-point of the cyan line segment, we can see that this policy attains at least $83\%$ of the optimal revenue). 

To gain some intuition behind the expressions of Theorem \ref{StaticGuaranteeMultiserverWithK}, we first note that the revenue guarantee is nothing but the complement of the blocking probability of an $M/M/C/\gamma+1$ queue when the arrival rate is $C$. This blocking probability decreases when more servers are added and also when the capacity $\gamma+1$ is increased, and so the opposite holds for our revenue guarantee: it improves with $C$ and $\gamma$. The cost guarantee, lacking a closed form except when $C=1$, is slightly harder to intuit. It is also increasing in $\gamma$, as can be seen in the $C=1$ case of Theorem \ref{StaticGuarantee}, but it is decreasing in the number of servers $C$. Put simply, our revenue and cost guarantees both improve with additional servers; on the other hand, with a higher cutoff $\gamma$, our revenue guarantee improves while cost guarantee gets worse.
 
We conjecture that the guarantees in Theorems~\ref{multiServerProfit} 
and~\ref{StaticGuaranteeMultiserverWithK} are tight, just as in the single‐server case. 
In fact, similar families of instances (e.g., those with linear demand, $\lambda = b - ap$, 
and appropriately chosen $a$ and $b$) appear to push each inequality in our proofs 
toward equality in multi‐server settings as well. However, identifying a closed‐form 
instance that rigorously demonstrates tightness is more challenging when $C>1$, 
due to the more complex state‐space and service parallelism. 

\subsection{Nearly tight examples}

We now provide some empirical support for our conjectured tightness of the profit, revenue, and cost guarantees.  First, consider the profit guarantee of Theorem~\ref{multiServerProfit}. We take the case $C=5$ with linear demand $\lambda = b - ap$, where $a = 10{,}000{,}000$ and $b = \left(1 + \tfrac{0.999}{C}\right)a$. In this instance, the optimal profit is $0.996264$, while the profit under the static cutoff policy arrival rate $\tilde\lambda$ and $\gamma=4$ is $0.7136287$, yielding a ratio of $0.7163$. This is quite close to the theoretical guarantee of $0.715$ furnished by Theorem~\ref{multiServerProfit}, lending credence to our belief that the bound is sharp.

Next, we examine the revenue guarantee of Theorem~\ref{StaticGuaranteeMultiserverWithK} in the same instance. The optimal revenue is $5.9835687$, while the static policy with $\gamma = 6$ achieves $4.9073757$, for a ratio of $0.8196$. This value is very close to the theoretical guarantee of $0.8185$. As $a \to \infty$, we expect this ratio to converge even more tightly to the bound. This experiment demonstrates that the simple linear-demand construction mirrors the single-server worst case and confirms the essential sharpness of our revenue guarantee.  

Finally, we illustrate the tightness of the cost guarantee of Theorem~\ref{StaticGuaranteeMultiserverWithK}. Here we consider $C=5$ with $a=66.7$, $b=73.4$, and cutoff $\gamma=6$. In this case, the optimal cost is $2.59288$, while the static policy yields $2.6188$. This gives a ratio of 1.0100285516, 
nearly the same as the theoretical guarantee $g(6,5) = 1.0100285581$. Together, these experiments lend credence to our conjecture profit, revenue, and cost guarantees are tight in the multi-server case as well.

\subsection{Proofs of Theorems \ref{multiServerProfit} and \ref{StaticGuaranteeMultiserverWithK}}

In the multi-server setting, the stationary probabilities are now expressed by the slightly altered expression
\begin{align}\label{multiServerSteadyState}
\mathbb{P}_i(\pi)
= \frac{\prod_{k=0}^{i-1} \frac{\lambda_k}{\mu_{k+1}} }{1+\sum_{n=0}^\infty \prod_{k=0}^n \frac{\lambda_k}{\mu_{k+1}} }
\end{align}
where $\mu_k = \min(C,k)$. By the same argument as before, we can again assume without loss of generality that $\mu=c=1$ for all proofs. Below, we extend Lemma \ref{singleserverstability} to the multi-server case and show that $\tilde{\lambda} < C$.
\begin{lemma}\label{multiServerArrivalRateBound} Suppose there are $C$ servers. Then for any policy $\pi = \{\lambda_0, \lambda_1, \dots\}$ which induces a stable queue,
$\tilde{\lambda} < C$.
\end{lemma}
\proof{Proof.}
We can write

\begin{align*}
\tilde{\lambda} 
&= \sum_{i=0}^\infty \lambda_i \mathbb{P}_i(\pi) \\
&= \sum_{i=0}^\infty \lambda_i
   \frac{\prod_{k=0}^{i-1} \tfrac{\lambda_k}{\mu_{k+1}}}
        {1+\sum_{n=0}^\infty \prod_{k=0}^n \tfrac{\lambda_k}{\mu_{k+1}}} \\
&= \frac{\sum_{i=0}^\infty \tfrac{C\lambda_i}{C}\prod_{k=0}^{i-1} \tfrac{\lambda_k}{\mu_{k+1}}}
        {1+\sum_{n=0}^\infty \prod_{k=0}^n \tfrac{\lambda_k}{\mu_{k+1}}} \\
&\leq \frac{C\sum_{i=0}^\infty \prod_{k=0}^i \tfrac{\lambda_k}{\mu_{k+1}}}
           {1+\sum_{n=0}^\infty \prod_{k=0}^n \tfrac{\lambda_k}{\mu_{k+1}}} \\
&< C.
\end{align*}

where the first inequality uses the fact that $\mu_i \leq C$ for all $i$ and the second  uses the fact that the denominator is finite because $\pi$ induces a stable queue.\
\Halmos \endproof

We now present the proof of our multi-server profit guarantees.
\proof{Proof of Theorem \ref{multiServerProfit}.}
We compute the revenue rate of $\pi^{\tilde\lambda, C-1}$ as
\begin{align*}
\mathcal{R}(\pi^{\tilde\lambda, C-1}) = \tilde\lambda p(\tilde\lambda) (1-\mathbb{P}_{C}(\pi^{\tilde\lambda, C-1}))\geq \mathcal{R}(\pi^*) (1-\mathbb{P}_{C}(\pi^{\tilde\lambda, C-1}))
\end{align*}
where the inequality follows from Eq. (\ref{RevBound}) which generalizes immediately to the multi-server case.

On the other hand, taking a look at the cost rate, we have
\begin{align*}
\mathcal{C}(\pi^{\tilde\lambda, C-1}) = \tilde\lambda(1-\mathbb{P}_{C}(\pi^{\tilde\lambda, C-1}))  \leq \mathcal{C}(\pi^*) (1-\mathbb{P}_{C}(\pi^{\tilde\lambda, C-1}))
\end{align*}
where the equality is Little's Law applied to the policy $\pi^{\tilde\lambda, C-1}$, and the inequality follows from Eq. (\ref{LittlesLaw1}), which also generalizes without modification to the multi-server case.

Thus we can arrive at 

\begin{equation*}
\begin{aligned}
\mathcal{Z}(\pi^{\tilde\lambda,\,C-1})
&= \mathcal{R}(\pi^{\tilde\lambda,\,C-1}) - \mathcal{C}(\pi^{\tilde\lambda,\,C-1}) \\
&\ge \mathcal{Z}(\pi^*)\bigl(1 - \mathbb{P}_C(\pi^{\tilde\lambda,\,C-1})\bigr) \\
&\ge \mathcal{Z}(\pi^*)\bigl(1 - \mathbb{P}_C(\pi^{C,\,C-1})\bigr). \quad \Halmos
\end{aligned}
\end{equation*}

\endproof
 
The following Lemma \ref{multiServerRev1} proves the claimed revenue guarantees of Theorem \ref{StaticGuaranteeMultiserverWithK}.
\begin{lemma}\label{multiServerRev1}
Suppose we have $C$ servers and $\pi^*$ is an optimal dynamic policy which induces the average arrival rate $\tilde\lambda$. Then for any $\gamma$ we have 

\begin{align*}
\Bigg(1-\frac{\frac{C^{\gamma+1}}{C!C^{\gamma+1-C}}}{\sum_{l=0}^C\frac{1}{l!}C^l + \frac{1}{C!}C^C (\gamma+1-C)}\Bigg)\mathcal{R}(\pi^*)\leq \mathcal{R}(\pi^{\tilde\lambda, \gamma})\end{align*}
\end{lemma}
\proof{Proof.}
As before, we can upper bound the revenue rate of the optimal policy via Jensen's
\begin{align}\label{multiServerJensen}
\mathcal{R}(\pi^*) \leq \tilde\lambda p(\tilde\lambda) 
\end{align}
Now we can simply compute the revenue rate of the policy $\pi^{\tilde\lambda, \gamma}$ using Eq. (\ref{revenueDefinition}) and proceed to arrive at a lower bound
\begin{align*}
\mathcal{R}(\pi^{\tilde\lambda, \gamma}) &= \tilde\lambda p(\tilde\lambda) \sum_{i=0}^{\gamma}\mathbb{P}_i(\pi^{\tilde\lambda,\gamma})\\ &= \tilde\lambda p(\tilde\lambda) (1-\mathbb{P}_{\gamma+1}(\pi^{\tilde\lambda,\gamma}))\\ &\geq \tilde\lambda p(\tilde\lambda) (1-\mathbb{P}_{\gamma+1}(\pi^{C,\gamma}))\\ &\geq (1-\mathbb{P}_{\gamma+1}(\pi^{C,\gamma}))\mathcal{R}(\pi^*)\\
&= \Bigg(1-\frac{\frac{C^{\gamma+1}}{C!C^{\gamma+1-C}}}{\sum_{l=0}^C\frac{1}{l!}C^l + \frac{1}{C!}C^C (\gamma+1-C)}\Bigg)\mathcal{R}(\pi^*)
\end{align*}
where the first inequality is from the fact that the blocking probability is increasing in $\tilde\lambda$ and the second inequality uses Eq. (\ref{multiServerJensen}). 
\Halmos \endproof

The following Lemma \ref{costLemmaMultiServer2} proves the claimed cost guarantee of Theorem \ref{StaticGuaranteeMultiserverWithK}.
\begin{lemma}\label{costLemmaMultiServer2}
Suppose we have C servers. If $\pi^*$ is the dynamic optimal policy, then the policy $\pi^{\tilde\lambda, \gamma}$ satisfies
\begin{align}
g(\gamma, C) \mathcal{C}(\pi^*)\geq  \mathcal{C}(\pi^{\tilde\lambda, \gamma})
\end{align}
where 
\begin{align*}
g(\gamma, C) = \max_{\tilde\lambda \in [0,C]}\frac{\sum_{i=1}^{\gamma+1}i\mathbb{P}_i(\pi^{\tilde\lambda, \gamma})}{\tilde\lambda}.
\end{align*}
\end{lemma}

\proof{Proof.}
First, we use Little's Law to bound the cost rate of the optimal dynamic policy \begin{align*}\mathcal{C}(\pi^*)  = \tilde\lambda E[W^*] \geq \tilde\lambda
\end{align*}
where we use the fact that $E[W^*] \geq 1$ because a sojourn always includes the service time.

On the other hand,
\begin{align*}
\mathcal{C}(\pi^{\tilde\lambda, \gamma}) = \sum_{i=1}^{\gamma+1}i\mathbb{P}_i(\pi^{\tilde\lambda, \gamma}) = \frac{\mathcal{C}(\pi^*)}{\mathcal{C}(\pi^*)}\sum_{i=1}^{\gamma+1}i\mathbb{P}_i(\pi^{\tilde\lambda, \gamma}) \leq \mathcal{C}(\pi^*)\frac{\sum_{i=1}^{\gamma+1}i\mathbb{P}_i(\pi^{\tilde\lambda, \gamma})}{\tilde\lambda}
\end{align*}

From Lemma \ref{multiServerArrivalRateBound}, we know that that the maximum value of the coefficient of $\mathcal{C}(\pi^*)$ is obtained for $\tilde\lambda \in [0,C]$. We can thus find the worst-case value and get a bound which does not depend upon $\tilde\lambda$. Indeed, letting $f(\gamma, C, \tilde\lambda) = \frac{\sum_{i=1}^{\gamma+1}i\mathbb{P}_i(\pi^{\tilde\lambda, \gamma})}{\tilde\lambda}$, we can write $g(\gamma, C) = \max_{\tilde\lambda \in [0,C]} f(\gamma, C, \tilde\lambda)$ so that we have
\begin{align*}
\mathcal{C}(\pi^{\tilde\lambda, \gamma}) \leq \mathcal{C}(\pi^*)f(\gamma, C, \tilde\lambda) \leq \mathcal{C}(\pi^*) g(\gamma,C). \Halmos
\end{align*}  \endproof

\section{Sojourn Time Penalty}\label{sojournModel}
Up to this point, the congestion penalty of a policy in our model has been given by the long-run average number of customers in the system under that policy. In this section, we analyze a new model where the congestion penalty of a policy is instead given by the long-run average sojourn times. As usual, the sojourn time is defined as the total time a customer spends in the system, including both the waiting time and the service time. Hence, this model is of practical importance as it relates directly to the quality of service experienced by the customers. 

The operation of this model is identical to the first. Potential customers arrive according to a Poisson process at rate $\Lambda$ and join the system if the offered price exceeds their valuation drawn i.i.d. from some regular distribution $F$. We let the subscript $s$ refer to the fact that this model uses sojourn time as the cost function. Letting $W(\pi)$ denote the steady-state sojourn time distribution of a customer who joins the system under policy $\pi$, the value rate $\mathcal{Z}_s(\pi)$ of a policy $\pi$ with rates $\lambda_0,\lambda_1,\ldots$ is now given by 
\begin{align*}
\mathcal{Z}_s(\pi) &= \sum_{i=0}^\infty \lambda_i p(\lambda) \mathbb{P}_i(\pi) - E[W(\pi)] \\
&= \sum_{i=0}^\infty p_i \lambda(p_i) \mathbb{P}_i(\pi) - \frac{E[L(\pi)]}{\sum_{i=0}^\infty \lambda_i \mathbb{P}_i(\pi) }\\
&= \sum_{i=0}^\infty \left(p_i \lambda(p_i)-\frac{i}{\sum_{i=0}^\infty \lambda_i \mathbb{P}_i(\pi) }\right) \mathbb{P}_i(\pi) 
\end{align*}
where the second equality uses Little's Law. Let $\pi^*_s$ denote the optimal policy for $\mathcal{Z}_s(\cdot)$ which has arrival rates $\lambda^*_{s,i}$. We use the notation $\tilde \lambda_s$ to denote the expected arrival rate under $\pi^*_s$, i.e, $\tilde \lambda_s=\sum_{i=0}^\infty \lambda^*_{s,i} \mathbb{P}_i(\pi^*_s)$.  
In this case, it is no longer possible to obtain guarantees on $\mathcal{Z}_s (\cdot)$ using static policies which pick the arrival rate $\tilde\lambda_s$. This fact is verified with an instance from the class described in Section \ref{tightnessExample}. If we take an instance with linear demand ($\lambda = b-ap$) with $b=6000$ and $a=5000$, we can observe that $\mathcal{Z}_s(\pi^*_s) \approx 0.17 > 0$, whereas $\mathcal{Z}_s(\pi^{\tilde\lambda_s, 0}) \approx -0.4 < 0$. For these instances, taking a cutoff higher than $0$ degrades performance further. Thus no result of the flavor of Theorem \ref{multiServerProfit} is possible, and hence we must resort to providing bi-criteria bounds on revenue and congestion in the spirit of Theorem \ref{StaticGuaranteeMultiserverWithK}.

 In fact, the situation here is worse still: guarantees for the class of static cutoff policies in general are not possible, not just for those which pick $\tilde\lambda$. Though for this class we can always attain a non-negative objective value ratio by taking the static policy which picks arrival rates of $0$, the performance can be arbitrarily bad. For example, for a single-server system with exponential demand, where $a=.463$ and $b=2$, the optimal static cutoff policy only attains $.1\%$ of the optimal objective value.

\begin{figure}[h]
\caption{A graphical representation of the guarantees of Theorem \ref{SojournGuarantee}}
\centering
\includegraphics[scale=.8]{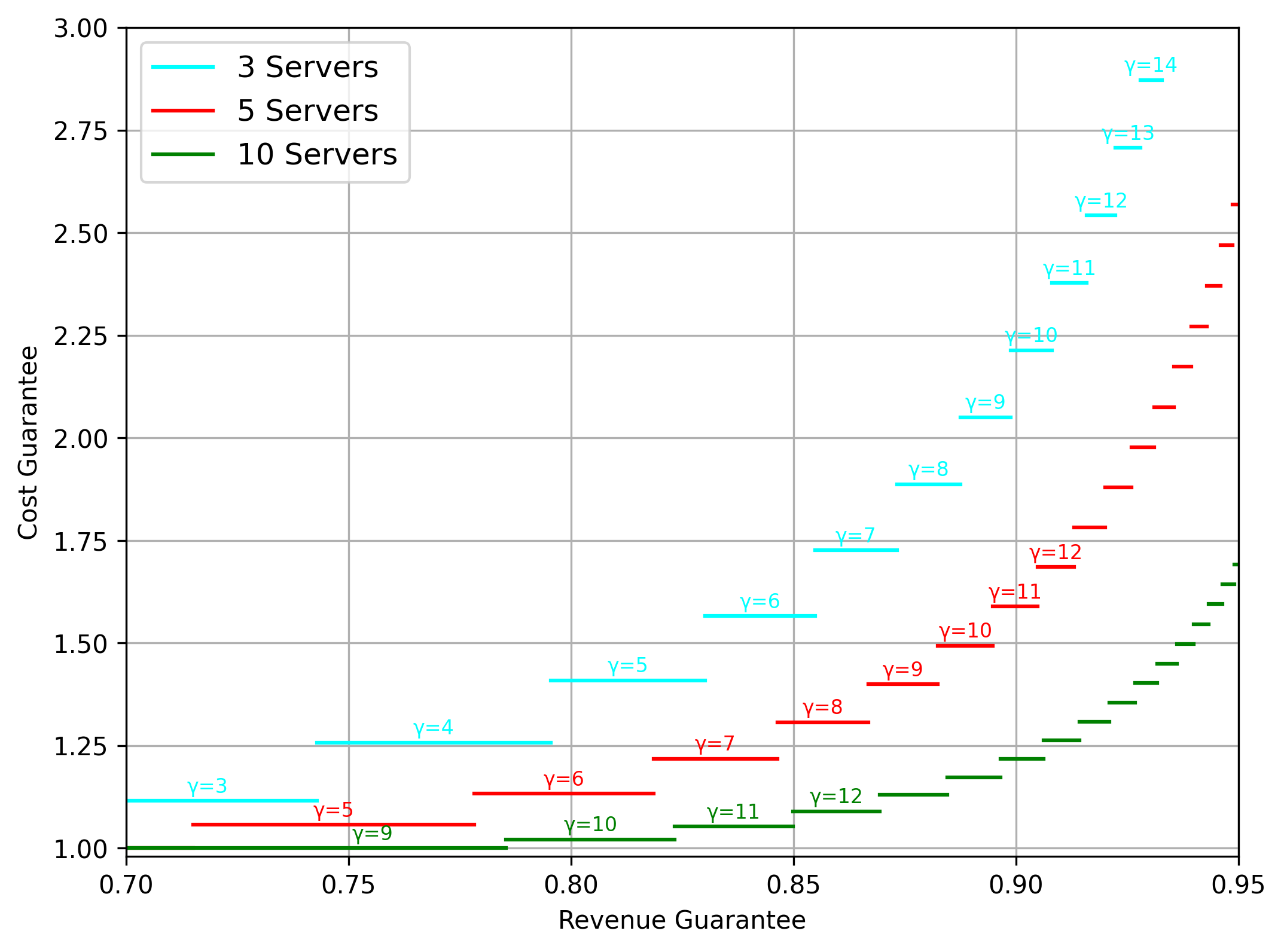} \label{fig2}



\vspace{.1cm}
\end{figure}

We present Theorem \ref{SojournGuarantee} below for the sojourn time penalty model for an arbitrary number of servers $C$. We observe that the congestion guarantees of the sojourn time model are not as strong as the guarantees we can furnish for the congestion model. Thus we are led to the insight that the control of sojourn times with static pricing is a marginally more difficult objective than just controlling congestion. The relative difference is most pronounced when looking at policies with small cutoff points, particularly $\gamma=1$ or $\gamma=2$.  In the single-server case, the congestion guarantee reduces to simply $ \frac{\gamma +2}{2}$ and the revenue guarantee reduces to $\frac{\gamma+1}{\gamma +2}$ for all $\gamma \geq 0$. 
We provide Figure \ref{fig2} to graphically illustrate the guarantees of Theorem \ref{SojournGuarantee} when we have $3, 5,$ or $10$ servers.
\begin{theorem}\label{SojournGuarantee}
For $\gamma \geq C-1$, the static pricing policy $\pi^{\tilde\lambda_s, \gamma}$ guarantees at least $1-\frac{\frac{C^{C}}{C!}}{\sum_{l=0}^C\frac{1}{l!}C^l + \frac{1}{C!}C^C (\gamma+1-C)}$ of the revenue rate and incurs a congestion penalty less than $\frac{\sum_{i=1}^C\frac{1}{(i-1)!}C^i + \frac{1}{C!}C^C\sum_{i=C+1}^{\gamma+1}i}{\sum_{l=0}^C\frac{1}{l!}C^{l+1}+\frac{\gamma-C}{C!}C^{C+1}}$ times that of the optimal dynamic pricing policy.
Equivalently,
\begin{align*}
    \Bigg(1-\frac{\frac{C^{C}}{C!}}{\sum_{l=0}^C\frac{1}{l!}C^l + \frac{1}{C!}C^C (\gamma+1-C)}\Bigg)\mathcal{R}(\pi^*_s)\leq \mathcal{R}(\pi^{\tilde\lambda_s, \gamma})
\end{align*}
and
\begin{align*}
    \frac{\sum_{i=1}^C\frac{1}{(i-1)!}C^i + \frac{1}{C!}C^C\sum_{i=C+1}^{\gamma+1}i}{\sum_{l=0}^C\frac{1}{l!}C^{l+1}+\frac{\gamma-C}{C!}C^{C+1}}E[W(\pi^*_s)]\geq E[W(\pi^{\tilde\lambda_s, \gamma})].
\end{align*}
\end{theorem}
\proof{Proof.}
As the revenue portion of our model remains unchanged, the revenue guarantee follows immediately from Theorem \ref{StaticGuaranteeMultiserverWithK}. Thus, it   suffices to prove the congestion guarantees.  
To remove the dependence on the value of $\tilde\lambda$, we seek the worst-case (largest) value of $E[W(\pi^{\tilde\lambda, \gamma})]$ over $\tilde\lambda \in [0,C]$. In contrast to the previous model, this coefficient of $\mathcal{C}(\pi^*)$ is non-decreasing in $\tilde\lambda$: clearly the expected sojourn time of a customer increases  if we increase the arrival rate to the system. Thus we can plug in $\tilde\lambda = C$ to get a universal bound.
\begin{align*}
E[W(\pi^{\tilde\lambda_s, \gamma})] 
&\leq E[W(\pi^*_s)]\,E[W(\pi^{\tilde\lambda_s, \gamma})] \\
&\leq E[W(\pi^*_s)]\,E[W(\pi^{C, \gamma})] \\
&= E[W(\pi^*_s)] \cdot 
   \frac{\sum_{i=1}^{\gamma+1} i \,\mathbb{P}_i(\pi^{C, \gamma})}
        {C\bigl(1-\mathbb{P}_{\gamma+1}(\pi^{C, \gamma})\bigr)}.
\end{align*}

where the first inequality follows since the expected sojourn time of any policy is at least the service time (which is assumed to be 1), the second is by the fact that sojourn times are non-decreasing in $\tilde\lambda$, and the equality uses Little's Law to compute $E[W(\pi^{C, \gamma})]$. Finally, we  use the expression for the multi-server steady-state probabilities in \eqref{multiServerSteadyState} to arrive at the claimed  guarantee.
\Halmos \endproof

\section{Numerical Experiments}\label{experimentsSection}

In this section, we present the results of numerical experiments on our static policies. Specifically, we consider linear ($\lambda = b-ap$), exponential ($\lambda = be^{-ap}$), and logistic ($\lambda = \frac{b\,(1+e^{-ap_0})}{1+e^{a(p-p_0)}}$) demand functions. We perform experiments with $1,3,5,$ and $10$‐server systems. Though these systems may be relatively small in the context of applications like food delivery and cloud computing, we will see that already with $C=10$ the performance is near-optimal, so further experiments would not be illustrative. For each number of servers $C \in \{1,3,5,10\}$, we sample $a$ uniformly from $[0.1,5]$, $b$ uniformly from $[.5,10]$, and $p_0$ uniformly from $[0,20]$. Under linear demand, when $a > b$, the optimal policy does not sell and thus achieves an objective value of $0$. Though this is also attainable by a static policy, for simplicity we reject these instances and resample. 

\noindent
To identify the optimal dynamic policy in each instance, we do \emph{not} rely on a direct Bellman‐type formulation, 
especially under the sojourn‐time penalty, as it is difficult to decompose sojourn costs into immediate rewards. 
Instead, we adopt a numerical approach that directly searches over all possible state‐dependent arrival rates:
\begin{enumerate}
    \item \textbf{Truncation:} For an $M$‐state queue, we choose $M$ large enough so that the stationary probability of being at state $M$ 
    is below a small threshold (e.g., 1\%). This ensures we capture essentially all of the probability mass.
    \item \textbf{Steady‐State Computation:} Given a candidate policy $(\lambda_0,\dots,\lambda_{M-1})$, we compute its steady‐state probabilities 
    (using the balance equations) and then evaluate the resulting revenue and congestion/sojourn costs.
    \item \textbf{SLSQP Optimization:} We optimize the  objective   with SciPy’s 
    \texttt{SLSQP} method.      Convergence is determined by default settings (objective changes below $10^{-6}$, iteration limits, etc.),    and we confirm these do not affect our final solutions. 
\end{enumerate}
This procedure reliably finds high‐quality policies across all tested instances. In simpler cases (small $M$) where 
we can verify an analytical or enumerative solution, the numerical method finds the correct optimum. 
Thus, the dynamic‐policy values reported in Tables \ref{tb:worst} and \ref{tb:avg} stem from this direct approach. We have made the code fully available at \url{https://github.com/jmb2457/StaticPricingforQueues}.

For each $C$ and each demand function, we perform 1000 replications and report the worst-case and average approximation ratios for the overall objective, revenue, and congestion penalties in Tables \ref{tb:worst} and \ref{tb:avg}.  
We report results for two different static policies. First, we consider the optimal static pricing policy. In other words, for each instance, we are simultaneously optimizing both the static price and the cutoff we choose. The notation $\lambda^*$ in the column headers refers to these policies. Second, we consider static policies which are constrained to use the price which induces the same average arrival rate as that of the optimal policy (these are the policies considered by our theorems). For each instance, we find the optimal dynamic policy, compute the average arrival rate, fix the price which induces this arrival rate, and then find the optimal cutoff for that fixed price. The notation $\tilde\lambda$ in the column headers refers to these policies.

\begin{table}[ht] 
\caption{Worst-case approximation ratios of static pricing policies with expected number in system penalty}\label{tb:worst}
\centering
\begin{tabular}{c|c||c|c||c|c||c|c}
\multicolumn{2}{c}{}   & \multicolumn{2}{c}{Objective}                 & \multicolumn{2}{c}{Revenue}     &            \multicolumn{2}{c}{Congestion}                \\ \hline
                             & $C$\rule{0pt}{2.5ex} & $\lambda^*$ & $\tilde\lambda$ & $\lambda^*$ & $\tilde\lambda$ & $\lambda^*$ & $\tilde\lambda$ \\ \hline
\multirow{4}{*}{Linear}      & 1\rule{0pt}{2.5ex}   & 93.0\%      & 75.7\%          & 80.8\%      & 69.1\%          & 111.6\%     & 114.0\%         \\ \cline{2-8} 
                             & 3\rule{0pt}{2.5ex}   & 95.4\%      & 95.1\%          & 92.1\%      & 87.8\%          & 109.5\%     & 119.0\%         \\ \cline{2-8} 
                             & 5\rule{0pt}{2.5ex}   & 98.0\%      & 97.9\%          & 97.2\%      & 97.2\%          & 108.3\%     & 121.0\%         \\ \cline{2-8} 
                             & 10\rule{0pt}{2.5ex}  & 99.9\%      & 99.9\%          & 99.9\%      & 99.9\%          & 100.0\%     & 100.0\%         \\ \hline
\multirow{4}{*}{Exponential} & 1\rule{0pt}{2.5ex}   & 89.7\%      & 89.0\%          & 82.6\%      & 75.7\%          & 105.4\%     & 119.0\%         \\ \cline{2-8} 
                             & 3\rule{0pt}{2.5ex}   & 97.4\%      & 97.1\%          & 96.6\%      & 98.5\%          & 103.2\%     & 117.0\%         \\ \cline{2-8} 
                             & 5\rule{0pt}{2.5ex}   & 99.6\%      & 99.6\%          & 99.5\%      & 99.9\%          & 101.2\%     & 103.0\%         \\ \cline{2-8} 
                             & 10\rule{0pt}{2.5ex}  & 99.9\%      & 99.9\%          & 99.9\%      & 99.9\%          & 99.9\%      & 100.0\%         \\ \hline
\multirow{4}{*}{Logistic}    & 1\rule{0pt}{2.5ex}   & 89.1\%      & 73.0\%          & 81.4\%      & 64.2\%          & 118.9\%     & 159.0\%         \\ \cline{2-8} 
                             & 3\rule{0pt}{2.5ex}   & 93.3\%      & 87.9\%          & 89.1\%      & 84.4\%          & 106.4\%     & 121.0\%         \\ \cline{2-8} 
                             & 5\rule{0pt}{2.5ex}   & 96.2\%      & 94.1\%          & 94.9\%      & 93.1\%          & 100.0\%     & 104.0\%         \\ \cline{2-8} 
                             & 10\rule{0pt}{2.5ex}  & 99.2\%      & 98.9\%          & 99.4\%      & 98.9\%          & 104.5\%     & 104.0\%      
\end{tabular}

\vspace{0.5cm}
\footnotesize \emph{Note.} The table compares optimal static pricing policies (attained by jointly optimizing both the rate $\lambda^*$ and the cutoff) and static policies which pick the rate induced by the optimal policy $\tilde\lambda$ (and then choose an optimal cutoff).
\end{table}
\begin{table}[ht] 
\caption{Average approximation ratios of static pricing policies with expected number in system penalty}\label{tb:avg}\centering
\begin{tabular}{c|c||c|c||c|c||c|c}
\multicolumn{2}{c}{}   & \multicolumn{2}{c}{Objective}                 & \multicolumn{2}{c}{Revenue}     &            \multicolumn{2}{c}{Congestion}                \\ \hline
                             & $C$\rule{0pt}{2.5ex} & $\lambda^*$ & $\tilde\lambda$ & $\lambda^*$ & $\tilde\lambda$ & $\lambda^*$ & $\tilde\lambda$ \\ \hline
\multirow{4}{*}{Linear}      & 1\rule{0pt}{2.5ex}   & 98.0\%      & 87.0\%          & 96.8\%      & 81.4\%          & 95.8\%     & 78.8\%         \\ \cline{2-8} 
                             & 3\rule{0pt}{2.5ex}   & 97.8\%      & 97.7\%          & 97.5\%      & 96.6\%          & 98.8\%     & 96.8\%         \\ \cline{2-8} 
                             & 5\rule{0pt}{2.5ex}   & 99.5\%      & 99.5\%          & 99.5\%      & 99.7\%          & 99.7\%     & 100.5\%         \\ \cline{2-8} 
                             & 10\rule{0pt}{2.5ex}  & 99.9\%      & 99.9\%          & 99.9\%      & 99.9\%          & 100.0\%     & 100.0\%         \\ \hline
\multirow{4}{*}{Exponential} & 1\rule{0pt}{2.5ex}   & 97.4\%      & 97.1\%          & 96.7\%      & 93.7\%          & 96.5\%     & 92.7\%         \\ \cline{2-8} 
                             & 3\rule{0pt}{2.5ex}   & 99.8\%      & 99.7\%          & 99.7\%      & 99.9\%          & 99.6\%     & 100.4\%         \\ \cline{2-8} 
                             & 5\rule{0pt}{2.5ex}   & 99.9\%      & 99.9\%          & 99.9\%      & 99.9\%          & 99.9\%     & 100.0\%         \\ \cline{2-8} 
                             & 10\rule{0pt}{2.5ex}  & 99.9\%      & 99.9\%          & 99.9\%      & 99.9\%          & 99.9\%      & 100.0\%         \\ \hline
\multirow{4}{*}{Logistic}    & 1\rule{0pt}{2.5ex}   & 96.0\%      & 85.2\%          & 95.6\%      & 86.7\%          & 95.0\%     & 99.0\%         \\ \cline{2-8} 
                             & 3\rule{0pt}{2.5ex}   & 97.8\%      & 94.6\%          & 98.1\%      & 95.3\%          & 99.6\%     & 101.0\%         \\ \cline{2-8} 
                             & 5\rule{0pt}{2.5ex}   & 96.2\%      & 94.1\%          & 94.9\%      & 93.1\%          & 100.0\%     & 104.0\%         \\ \cline{2-8} 
                             & 10\rule{0pt}{2.5ex}  & 99.9\%      & 99.9\%          & 99.9\%      & 99.9\%          & 100.5\%     & 100.3\%      
\end{tabular}

\vspace{0.5cm}
\footnotesize \emph{Note.} The table compares optimal static pricing policies (attained by jointly optimizing both the rate $\lambda^*$ and the cutoff) and static policies which pick the rate induced by the optimal policy $\tilde\lambda$ (and then choose an optimal cutoff).
\end{table}

On the whole, we see that static policies perform fairly well. Even in the worst-case instance found in our experiments, the static policy is able to attain $89.1\%$ of the optimal dynamic policy's objective value. Looking at the average performance, the difference becomes smaller still: even with only $1$ server, static policies lost at most 4\% on average compared to the optimal policy.

Moreover, the empirics echo our theoretical results: we see the advantage of dynamic over static pricing decreases rapidly as we increase the number of servers. For instances with $10$ or more servers, the performance of the optimal static policy is barely distinguishable from the dynamic optimum. This further lends credence to our practical insight that dynamic pricing may not be necessary for queueing systems with many servers.

The empirics also give some insight into how the static policies we construct in our proofs compare to the optimal static policies. The difference is sensitive to the form of the demand function. For exponential demand, the difference is fairly marginal: the largest discrepancy we observe in the worst-case objective ratios is $0.7\%$ when $C=1$. For linear and logistic demand functions, on the other hand, there are much larger discrepancies. In any case, the discrepancies diminish rapidly as we increase the number of servers. 

\subsection{Congestion Measured by Sojourn Time }
We also report the results of another set of experiments carried out in the same manner in Tables \ref{tb:worst_sojourn} and \ref{tb:avg_sojourn}, but this time for the model which takes the average sojourn time rather than the average number in system as the cost term. In particular, no objective value guarantees are possible because static cutoff policies can perform arbitrarily poorly in the sojourn model. Indeed, in the worst-case out of 1000 instances with one server and exponential demand, the optimal static cutoff policy was only able to recover $0.3\%$ of the optimal objective value. Moreover, when we use our construction and consider only static cutoff policies which pick the average arrival rate of the optimal dynamic policy, $\tilde\lambda_s$, in some cases we cannot even attain a positive objective value.

\begin{table}[ht]
\caption{Worst-case approximation ratios of static pricing policies with sojourn penalty}\label{tb:worst_sojourn}
\centering
\begin{tabular}{c|c||c|c||c|c||c|c}
\multicolumn{2}{c}{}   & \multicolumn{2}{c}{Objective}                 & \multicolumn{2}{c}{Revenue}     &            \multicolumn{2}{c}{Sojourn}                \\ \hline
                             & $C$\rule{0pt}{2.5ex} & $\lambda^*_s$ & $\tilde\lambda_s$ & $\lambda^*_s$ & $\tilde\lambda_s$ & $\lambda^*_s$ & $\tilde\lambda_s$ \\ \hline
\multirow{4}{*}{Linear}      & 1\rule{0pt}{2.5ex} & 34.8\% & -8169.0\% & 77.0\% & 63.3\% & 68.8\% & 73.8\% \\ \cline{2-8} 
                             & 3\rule{0pt}{2.5ex} & 2.3\% & -3.8\% & 92.9\% & 92.3\% & 95.7\% & 95.2\% \\ \cline{2-8} 
                             & 5\rule{0pt}{2.5ex} & 18.5\% & 17.1\% & 98.2\% & 98.3\% & 100.0\% & 100.0\% \\ \cline{2-8} 
                             & 10\rule{0pt}{2.5ex} & 100.0\% & 100.0\% & 100.0\% & 100.0\% & 100.0\% & 100.0\% \\ \hline
\multirow{4}{*}{Exponential}      & 1\rule{0pt}{2.5ex} & 0.3\% & -73.1\% & 64.3\% & 69.3\% & 60.1\% & 76.8\% \\ \cline{2-8} 
                             & 3\rule{0pt}{2.5ex} & 0.4\% & 0.4\% & 83.2\% & 91.8\% & 91.8\% & 97.4\% \\ \cline{2-8} 
                             & 5\rule{0pt}{2.5ex} & 8.6\% & 3.4\% & 98.1\% & 98.1\% & 100.0\% & 100.0\% \\ \cline{2-8} 
                             & 10\rule{0pt}{2.5ex} & 76.3\% & 76.2\% & 100.0\% & 100.0\% & 100.0\% & 100.0\% \\ \hline 
\multirow{4}{*}{Logistic}      & 1\rule{0pt}{2.5ex} & 64.4\% & -1368.9\% & 74.4\% & 63.9\% & 65.3\% & 77.3\% \\ \cline{2-8} 
                             & 3\rule{0pt}{2.5ex} & 46.5\% & 40.3\% & 91.1\% & 88.9\% & 93.4\% & 88.3\% \\ \cline{2-8} 
                             & 5\rule{0pt}{2.5ex} & 76.5\% & 75.7\% & 95.1\% & 95.3\% & 99.2\% & 99.3\% \\ \cline{2-8} 
                             & 10\rule{0pt}{2.5ex} & 99.6\% & 98.4\% & 99.8\% & 98.4\% & 100.0\% & 95.4\% \\
\end{tabular}

\vspace{0.5cm}
\footnotesize \emph{Note.} The table compares optimal static pricing policies (attained by jointly optimizing both the rate $\lambda^*_s$ and the cutoff) and static policies which pick the rate induced by the optimal policy $\tilde\lambda_s$ (and then choose an optimal cutoff).
\end{table}

\begin{table}[ht]
\caption{Average approximation ratios of static pricing policies with sojourn penalty}\label{tb:avg_sojourn}
\centering
\begin{tabular}{c|c||c|c||c|c||c|c}
\multicolumn{2}{c}{}   & \multicolumn{2}{c}{Objective}                 & \multicolumn{2}{c}{Revenue}     &            \multicolumn{2}{c}{Sojourn}                \\ \hline
                             & $C$\rule{0pt}{2.5ex} & $\lambda^*_s$ & $\tilde\lambda_s$ & $\lambda^*_s$ & $\tilde\lambda_s$ & $\lambda^*_s$ & $\tilde\lambda_s$ \\ \hline
\multirow{4}{*}{Linear}      & 1\rule{0pt}{2.5ex} & 93.5\% & 6.0\% & 92.5\% & 82.1\% & 93.7\% & 98.2\% \\ \cline{2-8} 
                             & 3\rule{0pt}{2.5ex} & 93.0\% & 92.8\% & 97.2\% & 97.1\% & 102.9\% & 103.3\% \\ \cline{2-8} 
                             & 5\rule{0pt}{2.5ex} & 98.7\% & 98.6\% & 99.9\% & 100.0\% & 103.0\% & 104.2\% \\ \cline{2-8} 
                             & 10\rule{0pt}{2.5ex} & 100.0\% & 100.0\% & 100.0\% & 100.0\% & 100.0\% & 100.0\% \\ \hline 
\multirow{4}{*}{Exponential}      & 1 & 72.6\% & 70.6\% & 86.7\% & 88.4\% & 96.7\% & 100.0\% \\ \cline{2-8} 
                             & 3\rule{0pt}{2.5ex} & 87.4\% & 88.3\% & 98.1\% & 98.2\% & 103.8\% & 104.9\% \\ \cline{2-8} 
                             & 5\rule{0pt}{2.5ex} & 96.2\% & 96.0\% & 99.9\% & 100.0\% & 101.4\% & 101.7\% \\ \cline{2-8} 
                             & 10\rule{0pt}{2.5ex} & 99.9\% & 99.9\% & 100.0\% & 100.0\% & 100.0\% & 100.0\% \\ \hline 
\multirow{4}{*}{Logistic}      & 1\rule{0pt}{2.5ex} & 95.0\% & 73.3\% & 95.4\% & 86.0\% & 95.9\% & 117.2\% \\ \cline{2-8} 
                             & 3\rule{0pt}{2.5ex} & 97.1\% & 95.0\% & 98.1\% & 96.7\% & 103.6\% & 112.9\% \\ \cline{2-8} 
                             & 5\rule{0pt}{2.5ex} & 98.8\% & 98.3\% & 99.2\% & 98.9\% & 103.9\% & 107.9\% \\ \cline{2-8} 
                             & 10\rule{0pt}{2.5ex} & 100.0\% & 100.0\% & 100.0\% & 100.0\% & 100.2\% & 100.1\% \\
\end{tabular}

\vspace{0.5cm}
\footnotesize \emph{Note.} The table compares optimal static pricing policies (attained by jointly optimizing both the rate $\lambda^*_s$ and the cutoff) and static policies which pick the rate induced by the optimal policy $\tilde\lambda_s$ (and then choose an optimal cutoff).
\end{table}

On average, however, static policies can still perform well, particularly with larger systems, echoing the results for the previous model. For 10 server systems, optimal static policies ($\lambda_s^*$) and  those which we constrain to pick the arrival rate $\tilde\lambda_s$ perform, on average, nearly identically to the optimal dynamic policies. Notably, there are still outliers (for example, an instance with exponential demand for which the optimal static policy was only able to attain $76.3\%$ of the optimal objective value). Overall, we can observe empirically that, with static policies, controlling sojourn times is more difficult than simply the queue length; in other words, for systems where keeping sojourn times low is the primary concern, dynamic policies have a slightly larger edge over static policies.

\section{Conclusion}\label{conclusion}
In this work, we have furnished universal performance guarantees for static pricing policies in the context of a general queueing model with price-sensitive customers. Our work yields non-asymptotic approximation ratios on revenue and congestion objectives simultaneously through a novel construction motivated by an application of Little's law.  We  observe  that the advantage of dynamic over static pricing decays rapidly as the number of servers increases. We also provide bi-criteria guarantees on the revenue and cost objectives, with closed-form guarantees in the M/M/1 setting. In applications where controlling sojourn times rather than congestion is the more salient objective, we provided similar guarantees on the performance of static policies. For the practitioner wondering if implementing dynamic pricing is worth the additional complexity and potential adversarial customer behavior it might induce, our results provide useful insights.

There are many possible avenues to extend our results. One promising direction of interest would be to relax the assumption that the service times are exponential. With general i.i.d. service times, the optimal pricing policy now depends not only on the number in system but also on the remaining service times of the customers in service. 
There are many subtleties that must be accounted for to prove similar  results, but   at the heart of our argument are two very robust tools: Little's Law and Jensen's inequality. One may also attempt to provide guarantees that are instance-dependent, using demand and service rate information to provide tighter inequalities. For instance, Eq. \eqref{LittlesLaw1} uses the simple fact that the total time in the system is at least the average service time, which can be loose in congested settings. Another interesting direction is to consider static pricing in queueing networks that represent vehicle sharing systems, where fair policies are of general concern \citep{elmachtoub2024fair}. 

Lastly, our theoretical results have focused on a particular construction of static policies for which we can provide good theoretical guarantees. We have shown that the worst-case objective value ratio for our construction is $\frac{1}{2}$ which is tight, but our empirical results have shown the best static policy had a guarantee of $0.891$. Improving the guarantee by considering other policies is thus another interesting open question.

\begingroup \parindent 0pt \parskip 0.0ex \def\enotesize{\normalsize} \theendnotes \endgroup

%
%
%

 \ACKNOWLEDGMENT{The authors acknowledge the support of NSF grant CMMI-1944428 and IIS-2147361.}


\bibliographystyle{informs2014} 
\bibliography{biblio.bib} 





\end{document}